\documentclass[11pt]{article}
\usepackage{epsfig,psfig,float,amssymb,latexsym}
\usepackage[notref,notcite]{}
\input{epsf}
\newcommand{\Ckzps}{C_{S}^{\bar{K}^0\!\!,\,p}}
\newcommand{\Ckzpp}{C_{P}^{\bar{K}^0\!\!,\,p}}
\newcommand{\Ckmps}{C_{S}^{K^-\!\!\!,\,p}}
\newcommand{\Ckmpp}{C_{P}^{K^-\!\!\!,\,p}}
\newcommand{\Ckzns}{C_{S}^{\bar{K}^0\!\!,\,n}}
\newcommand{\Ckznp}{C_{P}^{\bar{K}^0\!\!,\,n}}
\newcommand{\mpi}{\mu}
\newcommand{\NP}[1]{ Nucl.\ Phys.\ {\bf #1}}

\newcommand{\AN}[1]{Ann. Phys. NY {\bf #1}}

\newcommand{\PRL}[1]{ Phys.\ Rev.\ Lett.\ {\bf #1}}

\newcommand{\be}{\begin{equation}}
\newcommand{\ee}{\end{equation}}
\newcommand{\ba}{\begin{eqnarray}}
\newcommand{\ea}{\end{eqnarray}}

\newcommand{\nn}{\nonumber}

\begin{document}
\hfill{IFIC$-$01$-$0117}

\hfill{FTUV$-$01$-$0117}
\begin{center}
\vspace{0.5cm}
\huge{\bf{Meson exchange in the weak decay of}}

\vspace{0.4cm}
\huge{\bf{ $\Lambda$ hypernuclei
and the $\Gamma_n/\Gamma_p$ ratio}}
%{\Huge{\bf{ Meson exchange in weak decay of}}
%
%\vspace{0.4cm}
%\Huge{\bf{ hypernuclei and the
%$\Gamma_{p}/\Gamma_{n}$ ratio}}}\\
%\vspace{0.4cm}
%\end{center}

\end{center}
\vspace{.3cm}

\begin{center}
{\huge{D. Jido, E. Oset and J. E. Palomar}}
\end{center}

\begin{center}
{\small{\it Departamento de F\'{\i}sica Te\'orica e IFIC, Centro Mixto
Universidad de Valencia-CSIC,\\
46100 Burjassot (Valencia), Spain}}
\end{center}

\vspace{1cm}

%--------------0ld abstract----------------------------------------
%\begin{abstract}
%  We take an approach to the $\Lambda$ non-mesonic weak decay in nuclei based on the
%  exchange of mesons. The one pion and one kaon exchange are considered,
%  together with the exchange of two pions, either correlated, leading to an
%  important scalar-isoscalar exchange
%  ($\sigma$ like exchange) or uncorrelated
%  (box diagrams). The exchange of the two interacting pions in 
%   the scalar-isoscalar channel is done using techniques of chiral 
%   unitary theory which lead to a sigma pole for the $\pi \pi$ scattering 
%   amplitude
%   in the physical region. A drastic reduction of the $\Gamma_n/\Gamma_p$ ratio
%   is obtained which goes from values around 1/8 for the one pion exchange alone
%   to values around unity, which are compatible with all present experiments
%   within errors. We also calculate absolute rates for different nuclei and
%   using Landau-Migdal interaction parameters $g^{\prime}$ of the natural size 
%   induced
%   by short range correlations we obtain decay rates for different nuclei 
%   in good agreement with present data. 
%\end{abstract}

%--------------abstract from the proceedings of Torino-----------------
\begin{abstract}
  We take an approach to the $\Lambda$ non-mesonic weak decay 
  in nuclei based on the
  exchange of mesons. The one pion
   and one kaon exchange are considered,
  together with the exchange of two pions, either correlated, leading to an
  important scalar-isoscalar exchange
  ($\sigma$-like exchange), or uncorrelated
  (box diagrams). 
% For reviced version
Extra effects of omega exchange in the scalar-isoscalar channel are
also considered.
Constraints of chiral dynamics are used to generate these
  exchanges.  
A drastic reduction of the OPE results for the 
   $\Gamma_n/\Gamma_p$ ratio
   is obtained and the new results are compatible with all present experiments
   within errors. The absolute rates obtained for different nuclei are also in 
   good agreement with experiment.
\end{abstract}

 [Key Word] $\Lambda$ weak decay,  $\Gamma_n/\Gamma_p$ ratio, chiral unitary
 theory.

\section{Introduction}

  The problem of the $\Gamma_n/\Gamma_p$ ratio is the most persistent and
  serious problem related to the non-mesonic decay of $\Lambda$ hypernuclei.
  The problem lies in the large discrepancy between the theoretical ratio 
  provided by the one pion exchange model (OPE) and the experiment. While 
  certainly the OPE model is too simplified, the different attempts
  improving on the model for the non-mesonic decay have not been more
  successful. The OPE 
  model, using exclusively the parity conserving part of the weak $\Lambda$ 
  decay vertex $H_{\Lambda \pi N}$, leads to a   $\Gamma_n/\Gamma_p$ ratio of
  1/14 \cite{Oset:1990ey} in nuclear matter. The ratio is much influenced by 
  the antisymmetry of
  the two-nucleon wave functions and if one neglects crossed terms this ratio
  becomes 1/5. If in addition one includes the parity violating term, which is
  less important than the parity conserving one for the non-mesonic decay, the
  ratio changes  to about 1/8 \cite{PRB97}. These results are
  somewhat different in  \cite{Dubach:1996dg} where a ratio of about 1/5 is
  claimed for the $^{12}_\Lambda$C nucleus although a value of 1/11 is obtained
  in nuclear matter and the $^5_\Lambda$He nucleus. Results become worse when 
  short range correlations are
  taken into account and the values range from 1/16 in \cite{Takeuchi:1985yy},
   to 1/8 in \cite{Inoue:1996fs}, 1/14 in \cite{PRB97} and  1/20 in 
   \cite{Dubach:1996dg}, all of them for $^5_\Lambda$He. The ratios are
   somewhat larger for $^{12}_\Lambda$C, 1/10 in \cite{PRB97} and 
   1/5 in \cite{Dubach:1996dg}. 
   
       There are still some discrepancies as to the precise numbers but there is
  a systematic  agreement in the fact that they range from about 1/5 to 1/20 and
  that numbers of the order of unity, as experiments suggest, cannot be
  accommodated with the OPE model.
  
     Experimentally one has results for $^5_\Lambda$He from 
  \cite{Szymanski:1991ik}  with a ratio 0.93$\pm$0.5 and for $^{12}_\Lambda$C with
  ratios $1.33^{+1.12}_{-0.81}$ \cite{Szymanski:1991ik}, $1.87^{+0.91}_{-1.59}$ 
  \cite{Noumi:1995yd} and 0.70$\pm$0.30, 0.52$\pm$0.16 \cite{montwill}. More recent
  results for $^{12}_\Lambda$C are still quoted as preliminary 
  \cite{Outa:2000kj,hashimoto} but also range in values around unity with large errors.
  
      The large discrepancy of the OPE predictions with the experimental data
      has stimulated much theoretical work. One line of progress has been the
      extension of the one meson exchange model including the exchange of
      $\rho,\eta,K,\omega,K^*$ in  \cite{PRB97} and 
  \cite{Dubach:1996dg}. The results obtained are somewhat contradictory since
  while in \cite{Dubach:1996dg} values for the $\Gamma_n/\Gamma_p$ ratio
 around 0.83 are quoted for $^{12}_\Lambda$C, the number quoted in 
 \cite{PRB97} is 0.07. Also, in \cite{PRB97} the same ratio
 is obtained for $^5_\Lambda$He and $^{12}_\Lambda$C while in 
 \cite{Dubach:1996dg} the value of the ratio in $^{12}_\Lambda$C is about twice
 larger than for $^5_\Lambda$He (see \cite{Oset:1998xz} for a further discussion on
 this issue).
 
    Another line of progress has been the consideration of two pion exchange.
  An early attempt in \cite{Bando:1988pn} including N and $\Sigma$ intermediate states
  in a box diagram with two pions did not improve on the ratio and it made it
  actually slightly worse. However, in \cite{Shmatikov:1994up} the $\Delta$
  intermediate states were also considered leading to an increase of the 
   the $\Gamma_n/\Gamma_p$ ratio, although no numbers were given. A similar
   approach 
   was followed in \cite{Itonaga:1995jk,Shmatikov:1994sp} where the exchange of two
   interacting pions through the $\sigma$ resonance was considered and found to
   lead also to improved results in the $\Gamma_n/\Gamma_p$ ratio. Although
   there are still some differences in the works and results of 
   \cite{Itonaga:1995jk,Shmatikov:1994sp} (see \cite{Oset:1998xz} for details) they share
   the qualitative conclusion that the $\Gamma_n/\Gamma_p$ ratio increases when
   the $\sigma$ exchange is considered. In \cite{Itonaga:1995jk} the ratio goes from
   0.087 for only pion exchange to 0.14 when the correlated two pions in the 
   $\sigma$ channel (and also the $\rho$, which does not change much the ratio)
   are considered.
   
      A third line followed so far has been to take the quark model point of
  view. The origins of this line of work come from the pioneering work of 
 \cite{Cheung:1983sg}, where pion exchange was considered beyond a certain distance
  and quark degrees of freedom before. Two recent works follow this line 
  \cite{Inoue:1996fs,Maltman:1994da} although there are some discrepancies
  between them and some sign ambiguity that has been recently clarified using
  arguments of current algebra in \cite{Inoue:1998ep}. Quoting the results from
  this latter work, the $\Gamma_n/\Gamma_p$ ratio is changed from 0.13 for the
  OPE model to 0.49 when the quark degrees of freedom are considered in the
  nucleus of $^5_\Lambda$He. Further work is done in \cite{Oka:1999zi}, where
  considerations of chiral symmetry are done and interesting new relations are
  developed, but the conclusion is that, though the consideration of the quark
  degrees of freedom at short distances goes in the direction of improving
  the  $\Gamma_n/\Gamma_p$ ratio, the large contribution of the OPE makes the
  final ratios still incompatible with experimental results. However, some 
   recent
  advances in this direction, including $K$ exchange, \cite{Sasaki:2000vi,oka2} lead to
  improved ratios but also large widths.
  
    Some hopes were raised when the mechanism of the two-nucleon induced 
    $\Lambda$ decay was introduced in \cite{Alberico:1991wg} where the $\Lambda$
    decays into a nucleon and a virtual pion which is absorbed by two
    nucleons. The absorption of pions takes place mostly on neutron proton
    pairs, thus leading to a mechanism that produces two neutrons and a proton
    per $\Lambda$ decay. This enhances the production of neutrons versus protons
    and could be responsible for the large number of neutrons seen in the
    experiment without the need to have a large value for the 
    $\Gamma_n/\Gamma_p$ ratio of the one-nucleon induced $\Lambda$ decay.
     However, it was observed in \cite{Ramos:1994xy}
    that this had to be taken with care and, given the type of experimental
    analysis done to extract the $\Gamma_n/\Gamma_p$ ratio, the 
    consideration of this new mode made the experimental ratio even bigger,
     depending on the
    number obtained from the analysis without considering the two-nucleon mode.
    Actually, as noted in \cite{Gal:1995qe}, the results of the new analysis depend
    on whether the slow particles are detected or not. A detailed analysis
    of this problem considering final state interaction of the nucleons and
    actual detection thresholds was done in \cite{Ramos:1997ik} determining
    spectra of neutrons and protons from where future experiments can extract
    the $\Gamma_n/\Gamma_p$ ratio. For the purpose of the present paper the
    findings of \cite{Ramos:1997ik} simply tell us that the consideration of the
    two-nucleon mode of $\Lambda$ decay makes the experimental errors a little
    larger than assumed so far from present analyses.
  
     The situation is hence puzzling. Discrepancies between authors using a
    similar approach still persist, but in spite of that, there is a clear
    discrepancy between predictions of different models and present experimental
    results. 
    
        Our contribution to this problem has certainly benefitted from previous
    efforts and we have included in our approach in a unified way all the
    relevant elements
    considered before within the context of the one and two meson exchange.
    Thus,
    we include pion exchange, short range correlations via the Landau-Migdal
    interaction, which also serves to estimate the effects of the $\rho$ meson
    or $K^{*}$ exchange, kaon exchange, the exchange of two pions, 
     uncorrelated  or
    interacting in the scalar-isoscalar channel (the $\sigma$ channel) and omega
    exchange. The
    correlated two pion exchange has been done here following closely the steps
    of the recent work \cite{Oset:2000gn} where the two pions are allowed to
    interact, 
    using the Bethe-Salpeter equation and the chiral Lagrangians \cite{leu}.
    This chiral unitary approach to the $\pi \pi$ scattering problem leads to
    good agreement with the $\pi \pi$ data in the scalar sector including the
    generation of a pole in the t-matrix corresponding to the $\sigma$ meson
    \cite{Oller:1997ti}. We also study the $\omega$ exchange which is
    of the same order of magnitude.
    
     The results obtained here lead to ratios of $\Gamma_n/\Gamma_p$ of the 
     order of 0.53 and simultaneously one can obtain values for 
     the absolute rates
     of different nuclei that are comparable to the experimental ones 
     using reasonable Landau-Migdal parameters for the 
      strong p-wave interactions. These high values obtained for 
     $\Gamma_n/\Gamma_p$ are compatible with all  present experiments within
     errors, even more if these errors are enlarged as suggested in  
     \cite{Ramos:1997ik}.
     
       The present calculations are done in nuclear
     matter and the local density approximation is used to go to finite nuclei. The
     procedure should be quite good for nuclei around $^{12}_\Lambda$C and
     heavier, as done here. Yet, given the particular significance that the
     present findings have, apparently solving a long standing problem,
     additional calculations in finite nuclei should be encouraged and work in
     this direction is already under way \cite{assum}.
     
        The paper is organized as follows. In the next section we review the
	 OPE approach. In section 3 the kaon exchange is
	introduced. The two-pion exchange is included in section 4 and the
	results are presented and discussed in section 5. Conclusions are then
	presented in the last section.
	
\section{One Pion Exchange}
The  decay of the $\Lambda$ in nuclear matter is investigated here with
the propagator approach which provides a unified picture of different
decay channels of the $\Lambda$ \cite{Oset:1985js}. As mentioned above, the decay width of the $\Lambda$ is
calculated in  infinite nuclear matter, and is extended to finite
nuclei with the local density approximation. In this section we shall review the
calculation of the decay width of the $\Lambda$ in
nuclear matter using the one pion exchange approach.

First of all, we start with an effective $\pi \Lambda N$ weak interaction which is
written as, 
\begin{equation}
  {\cal L}_{\Lambda N \pi} = -i G \mpi^2 \bar{\psi}_N
   [ A +  \gamma_5 B] \vec{\tau}\cdot \vec{\phi}_\pi \psi_\Lambda +
   {\rm h.c.} 
\end{equation}
after the non-relativistic reduction we have:
\begin{equation}
  {\cal L}_{\Lambda N \pi} = -i G \mpi^2 \bar{\psi}_N 
   [S-(P/\mpi)\vec{\sigma}\cdot \vec{q}]
	\vec{\tau}\cdot \vec{\phi}_\pi \psi_\Lambda + {\rm h.c.}
  \label{lagpiLN}
\end{equation}
where $\mpi$ denotes the pion mass, and 
$G$ is the weak coupling constant with
\begin{center}
\begin{equation}
  G \mpi^2 = 2.211 \times 10^{-7}
\end{equation}
\end{center}
and $\vec{q}$ is the momentum of the outgoing pion.

By assuming that the $\Lambda$  behaves as a $I=1/2, I_z=-1/2$ state in the 
isospin space, 
this effective interaction already implements the phenomenological
$\Delta I = 1/2$ rule, which is seen in the nonleptonic free decay of 
the $\Lambda$. 
The coupling constants $S$ and $P$ are determined by the parity
conserving and parity violating amplitudes of the 
nonleptonic $\Lambda$ decay, respectively:
\begin{equation}
  A = S = 1.06, \hspace{1cm} {B \over 2M_N} \mpi =- P = - 0.527
\end{equation}
with $M_N$ the nucleon mass.
The $\pi NN$  vertex with strong interaction is given by the following
 effective Lagrangian:
\begin{equation}
   {\cal L}_{\pi NN}^S = - \frac{g}{2M_{N}} \bar{\psi}_N \gamma^{\mu}\gamma_5
    \vec{\tau}\cdot \partial_{\mu}\vec{\phi_\pi} \psi_N
\end{equation}
or, after the non-relativistic reduction,
\begin{equation}
   {\cal L}_{\pi NN}^S = -i {f_{\pi NN} \over \mpi} \bar{\psi}_N \vec{\sigma}
     \cdot \vec{q} \vec{\tau}\cdot \vec{\phi_\pi} \psi_N
\end{equation}
with $f_{\pi NN}=g\mpi/2M_N=1.02$, and an incoming pion of momentum $\vec{q}$.

\begin{figure}[ht]
\centering
\epsfysize=5.6cm
\epsfbox{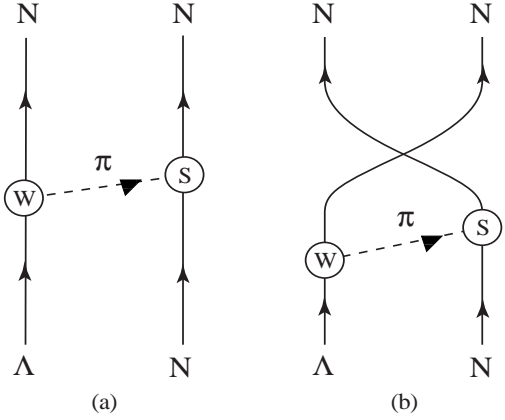}
\caption{Non-mesonic decay of $\Lambda$ with one $\pi$ exchange. 
  (a) and (b) denote the direct and exchange diagrams, respectively.}
  \label{fig1}
\end{figure}

\begin{figure}[ht]
\centering
\epsfysize=6.4cm
\epsfbox{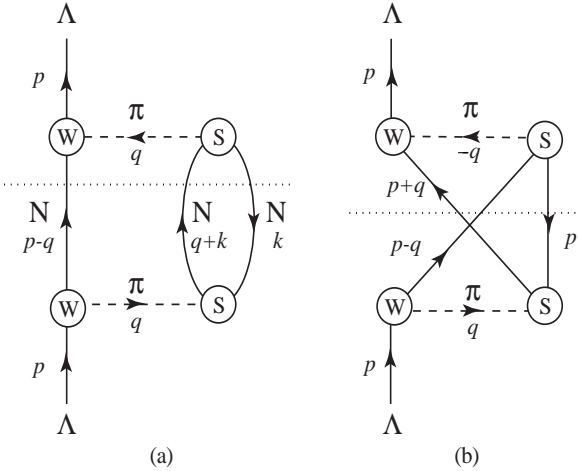}
\caption{Lowest order self-energy of the $\Lambda$. The cut gives
  the width of the $\Lambda$ for the corresponding non-mesonic decay
  of fig.~\ref{fig1}.}
  \label{fig2}
\end{figure}

% The non-mesonic decay of $\Lambda$ takes place with $\Lambda N
% \rightarrow NN$ transition in the nuclear medium, such as one pion
% exchange between $\Lambda$ and $N$ shown Fig.\ref{fig1}.
In order to evaluate the $\Lambda$ decay width $\Gamma$ in a nuclear
medium due to a certain $\Lambda N \rightarrow NN$ transition
amplitude, as depicted in fig.~\ref{fig1}, we start with the
calculation of the self-energy in the medium,  
$\Sigma$, shown in fig.~\ref{fig2}, 
and then we take its imaginary part:
\begin{equation}
  \Gamma = - 2\ {\rm  Im }\ \Sigma
\end{equation}
%%%% Figure 1 %%%%
% The medium effects are easily introduced into the
% calculation with modifications of the pion and nucleon propagations. 
%%%% Figure 2 %%%%

\begin{figure}[ht]
\centering
\epsfxsize=13cm
\epsfbox{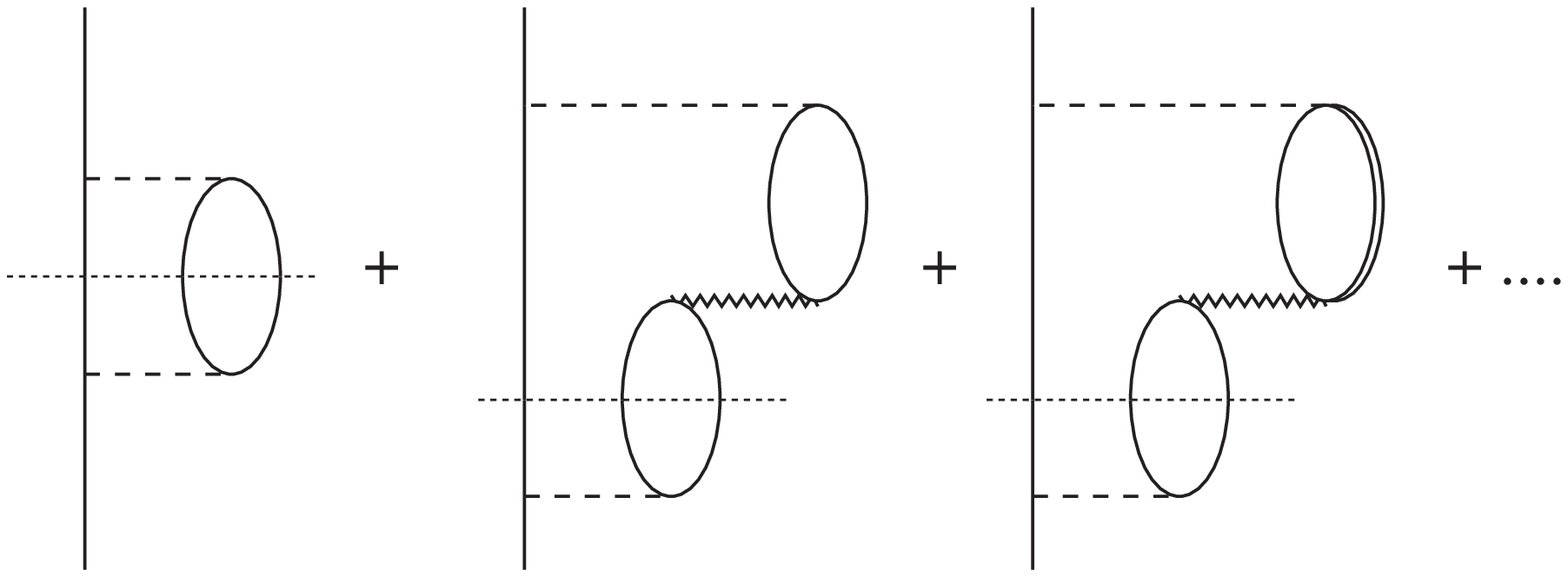}
\caption{The medium corrections to the $\Lambda$ self-energy shown in
fig.~\ref{fig2}.}
\label{fig3}  
\end{figure}

We take into account the $ph$ and $\Delta h$ excitations to all orders
in the sense of the random phase approximation (RPA), which is shown
in fig.~\ref{fig3}. 
%%%% Figure 3 %%%%
This induces modifications of the pion
propagation. 
% After the certain inclusion of the short range
% correlation (see appendix), we write the $NN$ effective interaction as
On the other hand, short range correlations modulate the $\Lambda N
\rightarrow NN$ transition amplitude in coordinate space and induce
changes in the momentum space representation (see Appendix). 
The p-wave (parity conserving) part of the weak $\Lambda N
\rightarrow NN$ transition is then written as
\begin{equation}
   G_{\Lambda N \rightarrow NN}^{\pi, {\rm p-wave}}(q) =
    [V_l^\prime(q)\hat{q}_i \hat{q}_j
   + V_t^\prime(q)(\delta_{ij}-\hat{q}_i \hat{q}_j)] 
    \sigma_i^{(1)} \sigma_j^{(2)} \vec{\tau}^{(1)} \cdot \vec{\tau}^{(2)} 
\end{equation}
with
\begin{eqnarray}
 V_l^\prime(q) &=& {f_{\pi NN} \over \mpi}{P \over \mpi} [\vec{q}^{\, 2} 
  D_\pi(q) F_\pi^2(q) + g_l^\Lambda(q)] \\
 V_t^\prime(q) &=& {f_{\pi NN} \over \mpi} {P \over \mpi} g_t^\Lambda(q) \ ,
\end{eqnarray}
%The explicit forms of the functions $ g_l^\Lambda(q)$ and
%$g_t^\Lambda(q)$ are given in the appendix. Here 
while the s-wave parity violating part gives rise to
\begin{equation}
   G_{\Lambda N \rightarrow NN}^{\pi, {\rm s-wave}}(q) =
   V_s^\prime(q)\ 
   \hat{q}_i \ \sigma_i^{(2)} \ \vec\tau^{(1)} \cdot \vec\tau^{(2)}  \ ,
\end{equation}
with
\begin{equation}
 V_s^\prime(q) = {f_{\pi NN} \over \mpi} S [
  D_\pi(q)F_\pi^2(q) +g_s^\Lambda(q)]|\vec{q}| 
\end{equation}
Here  $g_l^\Lambda(q)$, $g_t^\Lambda(q)$ and $g_s^\Lambda(q)$
implement the short range correlations, with a similar role
to that of the Landau-Migdal parameters, and 
$D_\pi(q)$ and $F_\pi(q)$ denote the pion propagator and form
factor, respectively.
 We use the same form factor $F_\pi(q)$ in both strong
and weak vertices:
\begin{equation}
    F_\pi(q) = { \Lambda_\pi^2 \over \Lambda_\pi^2 - q^2}
\end{equation} 

We have similar expressions for the spin-isospin strong effective interaction,
which we write as
\begin{equation}
G_{NN}(q) = [V_l(q)\hat{q}_i \hat{q}_j + V_t(q)(
    \delta_{ij}-\hat{q}_i \hat{q}_j)] \sigma_i^{(1)} \sigma_j^{(2)}
    \vec{\tau}^{(1)} \cdot \vec{\tau}^{(2)}
\end{equation}
with
\begin{eqnarray}
 V_l(q) &=& {f_{\pi NN}^{2} \over \mpi^2} [\vec{q}^{\, 2} D_\pi(q)F_\pi^2(q) + g_l(q)] \\
 V_t(q) &=& {f_{\pi NN}^{2} \over \mpi^2} [\vec{q}^{\, 2} D_\rho(q)F_\rho^2(q)
 C_\rho  + g_t(q)]
\end{eqnarray}
It is worth noting that the first $ph$ and $\Delta h$ excitations in fig.
\ref{fig3} are 
induced by the weak transition $G_{\Lambda N\rightarrow NN}(q)$, but,
once it happens, the successive excitations are produced by the strong
transition $G_{NN}(q)$ or the analogous one with $G_{N \Delta}(q)$.

The short range correlations, $g_l(q)$ and $g_t(q)$, which 
are written explicitly in eqs. (\ref{gl}, \ref{gt}) in the Appendix,
have slightly smaller size 
than those used in the conventional analysis of the spin-isospin
effective nuclear force \cite{NNforce}.  Thus we rescale these
functions $g_l(q)$ and $g_t(q)$
so that they have the value $g^\prime=0.7$ at $q=0$.

% Similarly the short range correlation split the p-wave effective
% interaction of $\Lambda N\rightarrow NN$ with weak interaction into
% the longitudinal and transverse components:
% Therefore  the transverse component gives the  smaller
% contribution than the longitudinal component, because 
% the transverse component is induced by $ph$ and $\Delta h$
% excitation and, hence, is suppressed at low density,
% unlike the effective $NN$ interaction, in which $\rho$
% meson can propagate.

In the nucleon propagator, the Pauli blocking effect is implemented in
terms of the nucleon occupation number $n(\vec k)$: 
\begin{equation}
   G_{N}(k) = {1-n(\vec{k}) \over k^0 - E(\vec{k})-V_N+i\epsilon}+
          {n(\vec{k}) \over k^0 - E(\vec{k})-V_N-i\epsilon}
\end{equation}
where the nucleon binding energy $V_N=-k_F^2/ 2M_{N}$ from 
the Thomas-Fermi approximation  and 
$n(\vec k)=1$ for $|\vec k|\leq k_F$, $n(\vec k)=0$ for 
$|\vec k| > k_F$ with $k_F$ the Fermi momentum, which depends on the
position of the $\Lambda$ through the density $\rho(r)$ in the local
density approximation. 

After the summation of the RPA series, the  
non-mesonic decay width coming from
the direct term, shown in fig.~\ref{fig1}(a), is obtained with the
result~\cite{Oset:1985js}:  
\begin{center}
\begin{displaymath}
\Gamma(k,\rho)=-(G\mpi^2)^2 \int {d^3 q \over (2\pi)^3}
    \theta(q_0)[ 1 -n(\vec{k}-\vec{q})] 
    {\rm \, Im\,} W(q)|_{q^0=k^0-E(\vec{k}-\vec{q})-V_N}
\end{displaymath}
\be
\label{gam}  
  {\rm \, Im\,} W(q) = {\rm \, Im\,}U_N(q) \ T(q) 
\ee
\end{center}
\noindent with:
\begin{center}
\begin{eqnarray}
\label{pidir} 
T^\pi_{p,{\rm dir}}(q) &=& \left[ 5 {V_s^{\prime 2}(q) \over |1-UV_l(q)|^2} +
          5 {V_l^{\prime 2}(q) \over |1-UV_l(q)|^2} +
         10 {V_t^{\prime 2}(q) \over |1-UV_t(q)|^2} \right]\ 
   \nn \\
   T^\pi_{n,{\rm dir}}(q) &=& \left[  {V_s^{\prime 2}(q) \over |1-UV_l(q)|^2} +
           {V_l^{\prime 2}(q) \over |1-UV_l(q)|^2} +
         2 {V_t^{\prime 2}(q) \over |1-UV_t(q)|^2} \right]\  
\end{eqnarray}
\end{center}
% \begin{equation}
%    T^\pi_{\rm dir}(q) = \left[ 6 {V_s^{\prime 2}(q) \over |1-UV_l(q)|^2} +
%           6 {V_l^{\prime 2}(q) \over |1-UV_l(q)|^2} +
%          12 {V_t^{\prime 2}(q) \over |1-UV_t(q)|^2} \right]\ ,
%   \label{pidir}
%  \end{equation}
$T^\pi_{p,{\rm dir}}$ and  $T^\pi_{n,{\rm dir}}$ together with eq. (\ref{gam}) give the $\Lambda$ decay induced by proton and neutron,
respectively.   
Here $U(q)=U_N(q)+U_\Delta(q)$, and $U_N$ and $U_\Delta$ are the
Lindhard functions for $ph$ and $\Delta h$ excitations.
To obtain eqs. (\ref{pidir}) we take the spin average of $\Lambda$
which removes the term linear in $\vec{\sigma}$ in the
$\Lambda$ self-energy. %The spin summation with the sign from the
%nucleon loop-* gives a factor $-1$ to
%both $V_s^{\prime 2}$ and $V_l^{\prime 2}$ while a factor $-2$ to
%$V_t^{\prime 2}$. The factor $6$ comes from  the isospin summation.

Up to here we have only considered the direct term. Actually 
the contribution from the exchange terms, shown in fig.~\ref{fig2}(b)
is suppressed by 
factor $1/2$ compared with the direct term, 
due to the absence of the nucleon
loop. Isospin indices will be
taken into account explicitly.
However, as we shall see later,
the exchange term gives a large contribution to the ratio of the
neutron induced decay to the proton induced one.

To calculate the exchange term shown as fig.~\ref{fig2}(b), we may use
the same set of propagator with the direct term.
In fact, assuming the $\Lambda$ at rest, the variable in the upper
propagator in fig.~\ref{fig2}(b) is $-\vec q-\vec p$ \ instead of $\vec
q$ in the direct diagram. However since $q$ is large ($\sim 420$ MeV/c)
and $p$ (the momentum of the occupied nucleons in the Fermi sea) is
smaller than $q$ and sometimes adds and other subtracts to $\vec q$, the
corrections are of order $(p/q)^2$, and hence $(p/q)^2$ is about $20\%$
in a term with smaller strength than the direct term and we neglect
$p$ in this interaction in front of $\vec q$.
%{\it *** should put here the reason and the estimation of error ***}. 
Therefore the difference in the calculation of the contribution from
the exchange terms is the summation of spin and isospin.
% The isospin summation gives a factor $-3$ instead of $6$ in the direct
% term.  
For the spin summation, it is important to
notice that the upper pion carries momentum $-q$, which produces a
different sign in the parity violating part with respect to the direct term. 
We calculate 
the spin sum of the exchange term for the parity violating part: 
\begin{equation}
 \frac{1}{2} \sum_s \langle s|\sigma^j\sigma^l|s \rangle 
  \hat{q}_j \hat{q}_l i(-iV_s^\prime)i(i V_s^\prime) =
  -V_s^{\prime 2}
\end{equation}

\noindent This gives the same sign as the direct term when including in the latter the
minus sign due to 
the fermion loop. The spin sum for the p-wave in the exchange term is given by:
\begin{eqnarray}
  \lefteqn{ \frac{1}{2} \sum_s \langle s|\sigma^i \sigma^k
\sigma^j \sigma^l |s \rangle} \nonumber \\
&& \ \ \times  \{V_l^\prime\hat{q}_i \hat{q}_j + 
    V_t^\prime(\delta_{ij}-\hat{q}_i \hat{q}_j)\}
    \{V_l^\prime\hat{q}_k\hat{q}_l + V_t^\prime(\delta_{kl}-\hat{q}_k
  \hat{q}_l)\}  \\
  &=& V_l^{\prime 2} - 4 V_l^\prime V_t^\prime
\end{eqnarray}
Note that $V_t^{\prime 2}$ vanishes but a crossed term between
$V_l^\prime$ and $V_t^\prime$ remains.

\begin{figure}[ht]
\centering
\epsfysize=8cm
\epsfbox{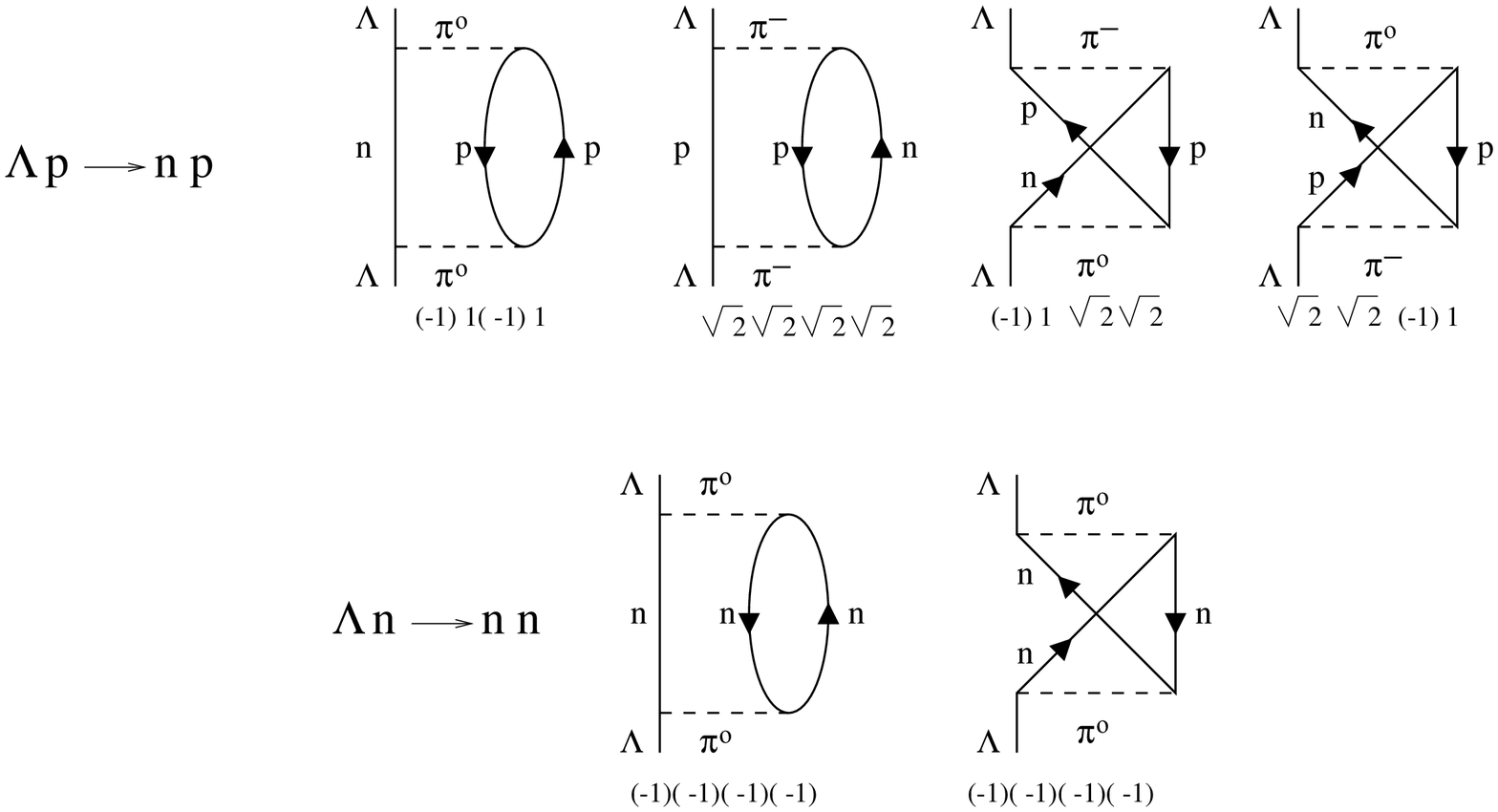}
\caption{The isospin factors of the direct and exchange terms induced
by proton and neutron. Recall that there is an extra relative minus sign for the
exchange terms because of the absence of a fermion loop with respect to the direct
term.}
\label{fig4}
\end{figure}

In fig.~\ref{fig4} one can see the different OPE diagrams corresponding to the
$\Lambda p\rightarrow n p$ and  $\Lambda n\rightarrow n n$ transitions. We
finally obtain:
\begin{equation}
   T^\pi_{p,{\rm exch}} = - {2 V_s^{\prime 2} \over |1-UV_l|^2} +
          {2 V_l^{\prime 2} \over |1-UV_l|^2} 
              -   8V_l^\prime V_t^\prime
         {\rm Re}\left[ {1 \over (1-UV_l)(1-U^*V_t)}\right] 
 \label{piexp}
\end{equation}
\begin{equation}
   T^\pi_{n,{\rm exch}} = {{1 \over 2}  V_s^{\prime 2} \over |1-UV_l|^2} -
          {{1 \over 2} V_l^{\prime 2} \over |1-UV_l|^2} 
              + 2 V_l^\prime V_t^\prime
         {\rm Re}\left[ {1 \over (1-UV_l)(1-U^*V_t)}\right] 
 \label{piexn}
\end{equation}
% \begin{equation}
%   T^\pi_{\rm exch} = -{3 \over 2} {V_s^{\prime 2} \over |1-UV_l|^2} +
%         {3 \over 2} {V_l^{\prime 2} \over |1-UV_l|^2} 
%              -   6V_l^\prime V_t^\prime
%         {\rm Re}\left[ {1 \over (1-UV_l)(1-U^*V_t)}\right] 
% \label{piex}
% \end{equation}
%%%% Figure 4 %%%%

With the direct and exchange terms we obtain the non-mesonic decay width
with one pion exchange. As we shall discuss later, if we add the mesonic width
and the two-nucleon induced width we obtain
$\Gamma_{tot}=1.6 \textrm{ }\Gamma^{free}_{\Lambda}$, 
which is larger than the experimental width.
If we neglect the transverse part and the induced interaction and we take 
only the direct terms, the ratio
$\Gamma_n /\Gamma_p$ is  $1/5$. However with the exchange terms,
it becomes
\begin{equation}
  {\Gamma_n \over \Gamma_p} = {{3 \over 2} \Gamma_S + {1 \over 2}
  \Gamma_P \over 3 \Gamma_S + 7 \Gamma_P} \label{Rpi}
\end{equation}
where $\Gamma_S$ and $\Gamma_P$ denote the partial width of
the parity violating and parity conserving decay. Eq. (\ref{Rpi}) gives 
$\Gamma_n /\Gamma_p = 1/2\sim 1/14$ (the values $1/2$ and $1/14$ are obtained 
when
considering only s-wave and p-wave respectively). Therefore the exchange terms
should be counted to reproduce both the total decay width and the $\Gamma_n
/\Gamma_p$
ratio. An actual calculation with all terms included gives $\Gamma \sim 1.1\textrm{
} 
\Gamma_{\Lambda}^{free}$
 and $\Gamma_n
/\Gamma_p \sim 1/8$ for $^{12}_{\ \Lambda}{\rm C}$.
This implies that we need some extra mechanisms additional to that of pion
exchange.

% Up to now we do not distinguish between the decays 
% of $\Lambda$ induced proton and neutron, but sum up these two contributions. 
% To get the ratio $\Gamma_n/\Gamma_p$, we separate the proton induced
% and neutron induced parts.
% For this purpose , we calculate
% the isospin factors in each diagrams, which is summarized in Fig.\ref{fig4}.
% Using that table we obtain 
% \begin{eqnarray}
%   T^\pi_p(p) &=& 5 V_s^{\prime 2} + 5 V_l^{\prime 2} + 10 V_t^{\prime
%    2} \nonumber  \\
%    && -2 V_s^{\prime 2} + 2 V_l^{\prime 2}  - 
%     8 V_l^\prime V_t^\prime
% \end{eqnarray}
% \begin{eqnarray}
%   T^\pi_n(p) &=&  V_s^{\prime 2} +  V_l^{\prime 2} + 
%     2 V_t^{\prime 2}\nonumber \\
%     && + {1 \over 2} V_s^{\prime 2} - {1 \over 2} V_l^{\prime 2}
%     + 2 V_l^\prime V_t^\prime
% \end{eqnarray}
% where the upper line comes from the direct terms, while the lower line
% from the exchange terms.
% Here we do not write the higher order of the RPA series, which,
% however, are easily recovered with putting the denominator as 
% (\ref{pidir},\ref{piex}). 

\section{Kaon Exchange}
\begin{figure}[ht]
\centering
\epsfysize=5.8cm
\epsfbox{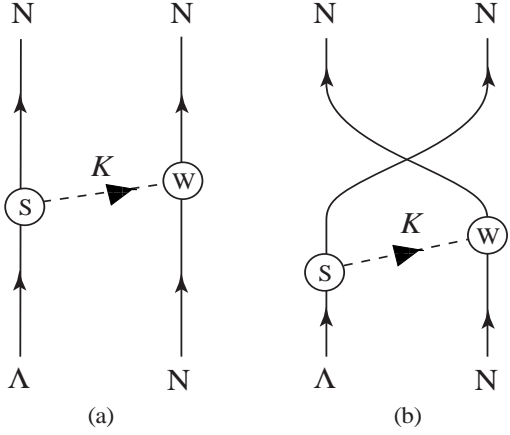}
\caption{Non-mesonic decay of the $\Lambda$ with one $K$ exchange. 
  (a) and (b) denote the direct and exchange diagrams, respectively.}
  \label{fig5}
\end{figure}

% There are some works on the non-mesonic decay of $\Lambda$ with pi
% and $K$ exchanges\cite{PRB97,SIO00}, which 
The $\Lambda$ non-mesonic decay with one $K$ exchange takes place
through the diagram shown in fig.~\ref{fig5}.
Including the $K$ exchange is  straightforward in the meson
propagator approach, once the $KNN$ weak vertex is fixed. 
%%%% Figure 5 %%%%

%%%% Figure 6 %%%%
\begin{figure}[ht]
\centering
\epsfysize=6.0cm
\epsfbox{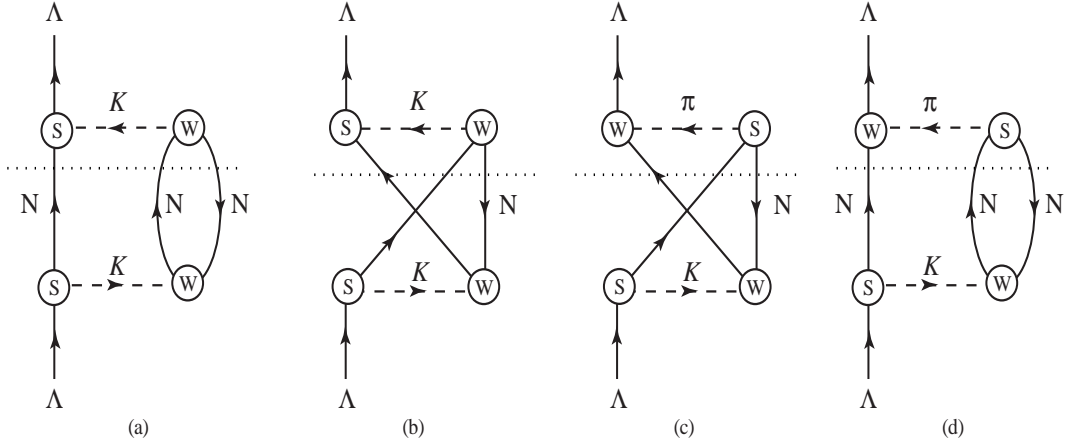}
\caption{Lowest order self-energy of the $\Lambda$. The cut gives
  the width of the $\Lambda$ for the corresponding non-mesonic decay
  to fig.~\ref{fig5}(a) and (b). The diagrams (c) and (d) are the interference
  terms between $\pi$ and $K$.}
  \label{fig6}
\end{figure}

The strong $K\Lambda N$ vertex is given by:
\begin{equation}
   {\cal L}_{K\Lambda N}^S = - \frac{g_{K\Lambda N}}{2M_N} \bar{\psi_{N}}
   \gamma^{\mu}\gamma_5 \partial_{\mu}\phi_{K} \psi_\Lambda + {\rm h.c.} 
\end{equation}
% after the non-relativistic reduction by
% \begin{equation}
%    {\cal L}_{K\Lambda N}^S = i {f_{K\Lambda N} \over \mpi} \bar{\psi}
%   \vec\sigma \cdot \vec q \phi_{K} \psi_\Lambda \ .
% \end{equation}
with $g_{K \Lambda N} \equiv f_{K\Lambda N} 2 M/ \mu$, which 
% The coupling constant $f_{K\Lambda N}$ 
is estimated with the $SU(3)$
flavor symmetry:
\begin{equation}
    \frac{f_{K\Lambda N}}{\mu} = -{D+3F \over 2 \sqrt{3} f}
\end{equation}
with $D+F=1.26$, $D-F=0.33$ and $f$ the pion decay constant, $f=93$ MeV. Note that there is a different sign in
$f_{K\Lambda N}$ with respect to the $pp\pi^0$ vertex, $f_{\pi NN}/\mu
=(D+F)/(2f)$.

The weak vertex of $NNK$ may be written as
\begin{eqnarray}
   {\cal L}_{KNN} &=& -i G\mpi^2 \left[ \bar{\psi}_p (A^{\bar{K}^0,p} +
   \gamma_5  B^{\bar{K}^0,p} )\phi_{K^0}^\dagger \psi_p  \right. \nonumber \\
   && + \bar{\psi}_n (A^{K^-,p} + \gamma_5
   B^{K^-,p})\phi_{K^+}^\dagger \psi_p  \\
   && + \left. \bar{\psi}_n (A^{\bar{K}^0,n} + \gamma_5
   B^{\bar{K}^0,n})\phi_{K^0}^\dagger \psi_n   \right] \ + {\rm h.c.} \nonumber 
\end{eqnarray}
% \begin{eqnarray}
%   {\cal L} &=& i G\mpi^2 \left[ \bar{\psi}_p (S^{K^0,p} -
%   (P^{K^0,p}/\mpi) \vec \sigma \cdot \vec q)\phi_{K^0} \psi_p 
%   \right. \nonumber \\
%   && + \bar{\psi}_p (S^{K^-,p} -
%   (P^{K^-,p}/\mpi) \vec \sigma \cdot \vec q)\phi_{K^-} \psi_n \\
%   && + \left. \bar{\psi}_n (S^{K^0,n} -
%   (P^{K^0,n}/\mpi) \vec \sigma \cdot \vec q)\phi_{K^0} \psi_n
%   \right] \nonumber 
% \end{eqnarray}

It is convenient to write these couplings in terms of $S$ and $P$ 
for the pion exchange, hence we introduce the
factor $C^{K,N}$ as
\begin{equation}
  A^{K,N} \equiv S^{K,N} = C^{K,N}_S S \hspace{1cm} 
  {B^{K,N} \over 2 M_N} \mpi \equiv C^{K,N}_P P
\end{equation}

\noindent To avoid confusion recall that in the case of the pion there was a minus sign
connecting $B$ and $P$ which we omit deliberately here to define the constants 
 $C^{K,N}_P$. Isospin symmetry gives one constraint \cite{Dubach:1996dg}: 
\begin{equation}
  \Ckzns=\Ckzps+\Ckmps \hspace{1cm}   \Ckznp=\Ckzpp+\Ckmpp
\end{equation}

The coupling constants $C^{K,N}$ should be determined by some
theoretical analysis,
because the $KNN$ vertex cannot be determined directly by the experiment. 
Here we take the result of the conventional analysis of \cite{Dubach:1996dg},
where for the parity violating part the amplitude is assumed to behave as the
$6^{th}$ component of the SU(3) generators, and for the parity conserving part
the pole model is used.
The values of $C^{K,N}$ are summarized in table~\ref{tab:CC}.
\begin{table}
\begin{center}
   \begin{tabular}{|ccc|}
   \hline 
         & PV & PC \\
   \hline
   $\pi$ & $S=1.05$ & $P=0.527$ \\
   K & $\Ckzps = 1.94$ & $\Ckzpp =  0.93$ \\
     & $\Ckmps = 0.76$ & $\Ckmpp = - 2.64$ \\
     & $\Ckzns = 2.70$ & $\Ckznp = - 1.72$ \\
   \hline 
   \end{tabular}
   \caption{Coupling constants in the weak
   interaction.    }
  \label{tab:CC}
\end{center}
\end{table}

Now we include one $K$ exchange in the previous result.
In the $K$ exchange the positions of weak and strong vertex are
opposite to the pion exchange. 
When introducing  the short range correlation in the $K$
exchange, one obtains the p-wave effective
interaction of $\Lambda N\rightarrow NN$:
\begin{equation}
   G_{\Lambda N \rightarrow NN}^{K, {\rm p-wave}}(q) =
    [V^{\prime}_{l, K}(q)\hat{q}_i \hat{q}_j
   + V^{\prime}_{t, K}(q)(\delta_{ij}-\hat{q}_i \hat{q}_j)] 
    \sigma_i^{(1)} \sigma_j^{(2)} C^{K,N}_P 
\end{equation}
with
\begin{eqnarray}
 V_{l,K}^\prime(q) &=& -{f_{K\Lambda N} \over \mpi}
     {P \over \mpi} [\vec{q}^{\, 2} D_K(q)F_K^2(q) + g_{l,K}^\Lambda(q)] \\
 V_{t,K}^\prime(q) &=& -{f_{K\Lambda N} \over \mpi} 
   {P \over \mpi} g_{t,K}^\Lambda(q)
\end{eqnarray}
and the parity violating part is written as
\begin{equation}
   G_{\Lambda N \rightarrow NN}^{K, {\rm s-wave}}(q) =
   V_{s,K}^\prime(q)\  \hat{q}_i\  \sigma_i^{(1)}\  C^{K,N}_S 
\end{equation}
with 
\begin{equation}
 V_{s,K}^\prime(q) = -{f_{K\Lambda N} \over \mpi} S [ 
  D_K(q)F_K^2(q) + g_{s,K}^\Lambda(q)]|\vec{q}| 
\end{equation}

Here $D_K(q)$ is the propagator for $K$ and the form factor for $K$ is
defined as 
\begin{equation}
  F_K(q) = {\Lambda_K^2 \over \Lambda_K^2 -q^2}
\end{equation}
with $\Lambda_K = \Lambda_\pi = 1.0$ GeV.

In order to introduce one $K$ exchange in the previous results, we replace the $\pi$
propagator by the $\pi +K$ with the following rules:
\begin{equation}
   \begin{array}{lcccccc}
{\rm proton}  & \pi^0 :&  -V^\prime_{l,t} & \longrightarrow & \pi^0+\bar{K}^0: &
     -V^\prime_{l,t} + C^{\bar{K}^0\!\!,p}_P V^\prime_{l,t,K} \\
  & \pi^- :&  2V^\prime_{l,t} & \longrightarrow & \pi^-+K^-: &
     2V^\prime_{l,t} + C^{K^-\!\!,p}_P V^\prime_{l,t,K} \\
{\rm neutron}  & \pi^0  :&  V^\prime_{l,t} & \longrightarrow & \pi^0+\bar{K}^0: &
     V^\prime_{l,t} + C^{\bar{K}^0\!\!,n}_P V^\prime_{l,t,K} 
  \end{array}
\end{equation}

\noindent while in s-wave part of the exchange we replace:
\begin{equation}
   \begin{array}{lcccccc}
{\rm proton}  & \pi^0 :&  -V^\prime_{s}\sigma^{(2)}_{i}\hat{q}_{i} &
 \longrightarrow & \pi^0+\bar{K}^0: &
     -V^\prime_{s}\sigma^{(2)}_{i}\hat{q}_{i} + C^{\bar{K}^0\!\!,p}_S
     V^\prime_{s,K}\sigma^{(1)}_{i}\hat{q}_{i} \\
  & \pi^- :&  2V^\prime_{s}\sigma^{(2)}_{i}\hat{q}_{i} & \longrightarrow &
   \pi^-+K^-: &
     2V^\prime_{s}\sigma^{(2)}_{i}\hat{q}_{i} + C^{K^-\!\!,p}_S 
     V^\prime_{s,K}\sigma^{(1)}_{i}\hat{q}_{i} \\
{\rm neutron}  & \pi^0  :&  V^\prime_{s} \sigma^{(2)}_{i}\hat{q}_{i} & 
\longrightarrow & \pi^0+\bar{K}^0: &
     V^\prime_{s} \sigma^{(2)}_{i}\hat{q}_{i}+ C^{\bar{K}^0\!\!,n}_S 
     V^\prime_{s,K}\sigma^{(1)}_{i}\hat{q}_{i} 
  \end{array}
\end{equation}

In the direct terms of s-wave (see fig. \ref{fig6} (d))the pion cannot interfere with $K$,
because the spin sum vanishes in the nucleon loop.

The results with one pion and one $K$ exchange in the lowest order of the
RPA series are the following ones:
%%%% proton %%%%
%%%% s-wave %%%%
\begin{eqnarray}
   T_p^{\pi+K,{\rm s-wave}}(q)  
   &=& 5 V_s^{\prime 2} + (V_{s,K}^\prime \Ckzps)^2 
  + (V_{s,K}^\prime \Ckmps)^2 \label{38} \\
  &&+   (-V_s^\prime + V_{s,K}^\prime \nonumber
   \Ckzps) (2 V_s^\prime + V_{s,K}^\prime \Ckmps) \nonumber  \\
% \end{eqnarray}
%%%% p-wave %%%%
% \begin{eqnarray}
 T_p^{\pi+K,{\rm p-wave}}(q)  &=& (-V_l^\prime + V_{l,K}^\prime \Ckzpp)^2 
    + (2V_l^\prime + V_{l,K}^\prime \Ckmpp)^2 \nonumber \\ 
  &&+ 2 \{ (-V_t^\prime + V_{t,K}^\prime \Ckzpp)^2 + (2V_t^\prime +
   V_{t,K}^\prime \Ckmpp)^2 \} \nonumber \\
  && - (-V_l^\prime+V_{l,K}^\prime \Ckzpp)
   (2V_l^\prime + V_{l,K}^\prime \Ckmpp)  \\
   &&+2 (-V_l^\prime+V_{l,K}^\prime \Ckzpp)
  (2V_t^\prime + V_{t,K}^\prime \Ckmpp)\nonumber  \\ 
   &&+ 2 (2V_l^\prime + V_{l,K}^\prime \Ckmpp) 
   (-V_t^\prime+V_{t,K}^\prime \Ckzpp) \nonumber \\
% \end{eqnarray}
%%%%% neutron %%%%
%%%%% s-wave  %%%%
% \begin{eqnarray}
 T_n^{\pi+K,{\rm s-wave}}(q)  &=&   V_s^{\prime 2} +(V_{s,K}^\prime \Ckzns)^2
 \nonumber \\
 && + {1\over 2} (V_s^\prime
   +V_{s,K}^\prime\Ckzns )^2  \\
% \end{eqnarray}
%%%% p-wave  %%%%
% \begin{eqnarray}
    T_n^{\pi+K,{\rm p-wave}}(q)&=&   (V_l^\prime +V_{l,K}^\prime\Ckznp )^2  
   +      2(V_t^\prime+V_{t,K}^\prime\Ckznp )^2 \nonumber \\
&&    -{1\over 2} (V_l^\prime +V_{l,K}^\prime\Ckznp )^2 \label{41} \\
 && +   2(V_l^\prime+V_{l,K}^\prime\Ckznp )(V_t^\prime
   +V_{t,K}^\prime\Ckznp ) \nonumber 
\end{eqnarray}

In order to include all orders of the RPA series, we just replace
\begin{eqnarray}
   V^{\prime 2}_{i,(1)} &\rightarrow & 
        {V^{\prime 2}_{i,(1)} \over |1-UV_i|^2} \\
   V^{\prime}_{l,(1)}V^\prime_{t,(2)} + V^{\prime}_{l,(2)}V^\prime_{t,(1)}
    &\rightarrow & {\rm Re}\left\{ {1\over 1-UV_l} {1\over 1-U^*V_t}
   \right\}
   (V^{\prime}_{l,(1)}V^\prime_{t,(2)} + 
   V^{\prime}_{l,(2)}V^\prime_{t,(1)}) \nn
\end{eqnarray}
% \begin{eqnarray}
%   V^{\prime 2}_{i} &\rightarrow & {V^{\prime 2}_{i} \over |1-UV_i|^2} \\
%   V^{\prime 2}_{i,K} &\rightarrow & {V^{\prime 2}_{i,K} \over |1-UV_i|^2} \\
%   V^{\prime}_{l}V^\prime_{t,K} + V^{\prime}_{l,K}V^\prime_{t}
%    &\rightarrow & {\rm Re}\left\{ {1\over 1-UV_l} {1\over 1-U^*V_t}
%   \right\}
%   (V^{\prime}_{l}V^\prime_{t,K} + V^{\prime}_{l,K}V^\prime_{t})
% \end{eqnarray}

Here $V_{i,(1)}^\prime$ and $V_{i,(2)}^\prime$ with $i=l,t,s$ stand
for the effective interactions with $\pi$ or $K$ exchange.
The s-wave component of $K$, however,  remains $V^{\prime
2}_{s,K}$, because the vertex in the $Kph$ excitation contains no
$\vec \sigma$ and hence this $ph$ excitation does not get modified by the
effective strong spin-isospin interaction.
% it can induce only one $ph$ excitation due to the
% absence of spin sum of the nucleon loop.

\section{Two-pion exchange}

Another kind of diagrams that have been traditionally studied are those
corresponding to two-pion exchange. We will divide the study of these diagrams
into two categories: correlated two-pion exchange and uncorrelated two-pion
exchange. In the case of correlated exchange we will only consider the scalar-isoscalar channel, where the $\sigma$ meson appears. The vector channel is
neglected since the $\rho$ contribution has been seen to be not 
relevant~\cite{Shmatikov:1994sp,Parreno:pc}. Simple estimations will also be done here. 
 We will also see that the scalar-isoscalar channel is also the relevant one
  in the case of uncorrelated two-pion exchange.
 The effect of heavier scalar mesons (such as the $f_{0}, a_{0}$) is also 
  negligible~\cite{Shmatikov:1994sp}.

\subsection{Correlated two-pion exchange}

Some works on this topic have been done~\cite{Itonaga:1995jk,Shmatikov:1994sp,Itonaga:1998ua},
 incorporating
the $\sigma$ meson as an explicit degree of freedom. There it is found that,
working with reasonable values for the mass, width and $\sigma \pi \pi$
coupling, the role of this $"2\pi /\sigma"$
exchange is relevant in the non-mesonic decay problem.

A less phenomenological treatment of the sigma meson is provided by Chiral
Unitary Approaches~\cite{ Oller:1997ti,Dobado:1997ps, Oller:1998ng,
Oller:1999zr, Markushin:2000fa}. In~\cite{Oller:1997ti} it was found that the
$\sigma$ meson is dynamically generated by the interaction of 
two pion in flight when summing up the s-wave $t-$matrix  of the
$\pi \pi$ scattering to all orders using the Bethe-Salpeter equation. The former
 picture of the $\sigma$ meson was used to describe its role in the $NN$
interaction in ref.~\cite{Oset:2000gn}, finding a moderate attraction beyond $r=0.9$ 
fm and a repulsion at shorter distances, in contrast with the attraction of 
 the conventional $\sigma$ exchange. We will follow an analogous model to the one of
the aforementioned reference.

\begin{figure}[ht]
\centerline{\includegraphics[width=0.4\textwidth]{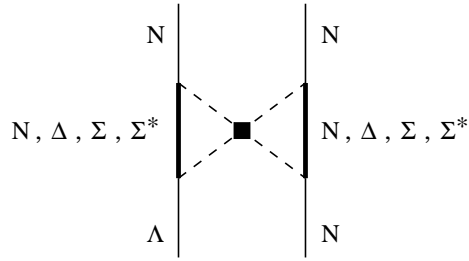}}
\caption{Diagrams corresponding to the correlated two-pion exchange, with
in-flight interaction of the two mesons.}
\label{Ndss}
\end{figure}

\begin{figure}[ht]
\centerline{\includegraphics[width=0.8\textwidth]{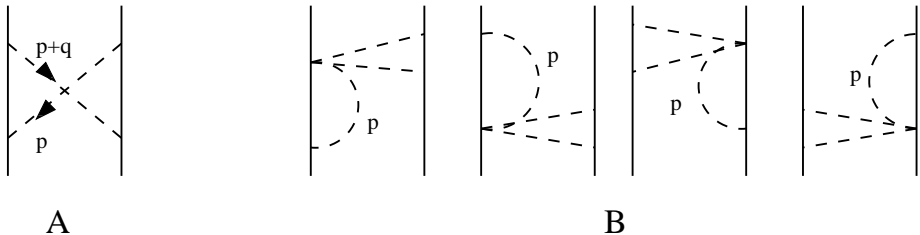}}
\caption{Diagrams where the off-shell cancellations appear. The
 off-shell part of the pion lines in diagram A cancels diagrams B.}
\label{cancel}
\end{figure}

The diagrams corresponding to the correlated exchange are those of 
 fig.~(\ref{Ndss}). In the weak vertex we will only consider at the moment the 
 parity conserving term of the Lagrangians (we will see later that the contribution from the parity violating part is not relevant).
This simplifies the problem because, as
 the parity conserving part (proportional to $\vec{\sigma}
 \vec{q}$, where $\vec{q}$ is the momentum of the pion) has the same structure
 as the strong $\pi NN$ vertex, the results obtained in ref.~\cite{Oset:2000gn}  are also
 applicable here. In that reference it was shown that there is a cancellation of
 the meson-meson vertex off-shell part of the diagram \ref{cancel}{A} with the diagrams
 \ref{cancel}{B}
  containing a $\pi \pi \pi NN$ vertex. 
Taking this fact into account, the rescattering of the pions can be treated
non-perturbatively by means of the Bethe-Salpeter equation, as was done in
ref.~\cite{Oller:1997ti}. Such a treatment also applies in the case of the $\Lambda N \rightarrow NN$ interaction when
considering only the parity conserving term of the 
interacting Lagrangian. One is then allowed to take the expression of the potential coming
from the diagrams with $N$ and $\Delta$ as intermediate states from that
reference, with the only
difference of a multiplicative factor $\mathfrak{R}$ that reflects the replacement of one strong
 $\pi NN$ vertex by the weak $\Lambda \pi N$. The potential is then given
 by~\cite{Oset:2000gn}:
 
 \begin{center}
 \begin{equation}
 \label{tt}
 V^{2\pi}_{corr}(q)=\mathfrak{R} \ v^{2}(q) \frac{6}{f^{2}}
\frac{\vec{q}^{\ 2}+\frac{\mu^{2}}{2}}{1-G(-\vec{q}^{\ 2})\left(\vec{q}^{\
2}+\frac{\mu^{2}}{2}\right)\frac{1}{f^{2}}}
\ee
\end{center}

\noindent where $G(s)$ is the loop function with two pion propagators, and
$v(q)$ (associated to diagrams of fig. \ref{Ndss} with intermediate N, $\Delta$)
 and $\mathfrak{R}$ are given by:

\begin{center}
\ba
\label{rv}
\mathfrak{R} &=& \frac{P/\mu}{\frac{D+F}{2f}}\nn\\
v(q) &=& v_{N}(q)+v_{\Delta}(q)
\ea
\end{center}

In the former equation $v_{N}(q)$ can be written, after performing
the $p^{0}$ integration and assuming that the $\Lambda$ baryon is initially at
rest, as:

\begin{center}
\ba
\label{vn}
v_{N}(q)=\int
\frac{d^{3}p}{(2\pi)^{3}}\left(\frac{D+F}{2f}\right)^{2}
(\vec{p}^{\ 2}+ \vec{p} \cdot \vec{q}) \frac{M_{N}}{2 
E_{N}}\frac{1}{\omega}\frac{1}{\omega'}\theta(|\vec{p}|-k_{F})\frac{1}{\omega+\omega'}
\nn\\
\times \frac{1}{E_{N}+\omega-M_{\Lambda}}\ \frac{1}{E_{N}+\omega'-M_{\Lambda}}(\omega+\omega'
+E_{N}-M_{\Lambda})
\ea
\end{center}

In the former equations the momenta $\vec{p}$ and $\vec{q}$ are those of fig.~(\ref{cancel}) and:

\begin{center}
\be
\label{eomegas}
E=E(\vec{p}) ; \ \  \ \ \ 
\omega=\sqrt{\mu^{2}+\vec{p}^{\ 2}};
\ \ \ \ \ \ \omega'=\sqrt{\mu^{2}+(\vec{p}+\vec{q})^{\ 2}}.
\ee
\end{center}

\noindent $\theta$ is the Heaviside function and $k_{F}$ is the Fermi
momentum. In eq. (\ref{vn}) we have neglected the energy transfer $q^0$ which is
small compared with $\vec{q}$.

The $v_{\Delta}(q)$ function has the same structure as
$v_{N}(q)$. The only difference is that we have to replace the
energy and the mass of the intermediate nucleon by those of the $\Delta$, and the
$\pi NN$ vertex by the $\pi N \Delta$ one. The Heaviside function disappears now
from the integrand since the $\Delta$ resonance is not Pauli-blocked, and there
is also an extra isospin coefficient:

\begin{center}
\ba
\label{vd}
v_{\Delta}(q)=\frac{4}{9} \left(\frac{f_{\pi
N\Delta}}{f_{\pi NN}}\right)^{2}\left(\frac{D+F}{2f}\right)^{2} \int
\frac{d^{3}p}{(2\pi)^{3}}
(\vec{p}^{\ 2}+ \vec{p} \cdot \vec{q}) \frac{M_{\Delta}}{2 
E_{\Delta}}\frac{1}{\omega}\frac{1}{\omega'}\frac{1}{\omega+\omega'}
\nn\\
\times \frac{1}{E_{\Delta}+\omega-M_{\Lambda}}\ \frac{1}{E_{\Delta}+\omega'-M_{\Lambda}}(\omega+\omega'
+E_{\Delta}-M_{\Lambda})
\ea
\end{center}

Here $f_{\pi N \Delta}$ parameterizes the $\pi N \Delta$ coupling, with $f_{\pi
N\Delta}=2.12 \textrm{ } f_{\pi NN}$.
The regularization of these loops can be accomplished by means of a form factor. We take 
 static form factors as in~\cite{Oset:2000gn}:
\begin{center}
\be
\label{ff}
F(\vec{p})F(\vec{p}+\vec{q})=\frac{\Lambda_{\pi}^{2}}{\Lambda_{\pi}^{2}+\vec{p}^{\ 2}}
\ \frac{\Lambda_{\pi}^{2}}{\Lambda_{\pi}^{2}+(\vec{p}+\vec{q})^{2}}\ \ ,
\ee
\end{center}

\noindent and in our calculations we take  $\Lambda_{\pi}$=1 GeV.

\begin{figure}[ht]
\centerline{\includegraphics[width=0.5\textwidth]{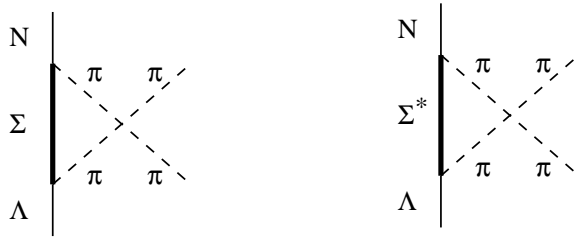}}
\caption{The two pion correlated exchange triangle vertex through $\Sigma$ and
$\Sigma^{*}$ excitations.}
\label{cancesig}
\end{figure}

\begin{figure}[ht]
\centerline{\includegraphics[width=0.4\textwidth]{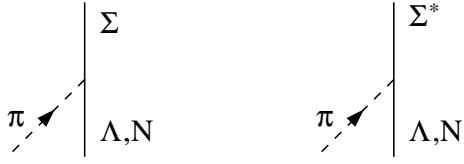}}
\caption{ Vertices needed to evaluate the $v_{\Sigma}(q)$ and
$v_{\Sigma^{*}}(q)$ functions.}
\label{vertex}
\end{figure}

So far we have been studying diagrams in which the $\Lambda$ baryon
appears in the weak vertices. However, as we can see in fig.~\ref{Ndss}, there
 are also diagrams with a strange intermediate baryon ($\Sigma, \Sigma^{*})$
 strongly 
 coupled to the $\Lambda$. In
 these new diagrams the cancellation of the  meson-meson
 vertex off-shell part of diagram \ref{cancel}{A} with the diagrams
 \ref{cancel}{B} also holds. Therefore,
  we only need to calculate the $v_{\Sigma}(q)$ and
 $v_{\Sigma^{*}}(q)$ functions (analogous to $v_{N}(q)$ and
 $v_{\Delta}(q)$) which correspond to the diagrams depicted in
 fig.~\ref{cancesig}. In order to calculate these functions we need to know the $\Lambda
 \Sigma \pi$, $\Lambda \Sigma^{*} \pi$, $\Sigma N \pi$ and $\Sigma^{*} N \pi$
 vertices (see fig.~\ref{vertex}).
   
%\begin{figure}[ht]
%\label{vertex}
%\centerline{\includegraphics[width=0.5\textwidth]{vertex.ps}}
%\caption{Vertex needed to evaluate the $V_{\Sigma}(q)$ and
%$V_{\Sigma^{*}}(q)$ functions.}
%\end{figure}
 
  The strong $\Lambda \Sigma \pi$ vertex is given in ref.~\cite{Nacher:2000vg}
 and the $\Lambda \Sigma^{*} \pi$ vertex can be obtained from the former one by
 replacing the  spin and isospin Pauli matrices by the analogous
 representation-changing operators $\vec{S}$ and $\vec{T}$ and by invoking
 $SU(6)$ symmetry, as done in~\cite{Oset:2001eg}.
  We can write the strong $\Lambda \Sigma \pi$ and $\Lambda \Sigma^{*} \pi$
  vertices as:
  
\begin{center}
\ba
\label{vertexsigstrong}
-it_{\Lambda \Sigma \pi}=\frac{1}{\sqrt{3}}\frac{D}{f} \vec{\sigma} \cdot \vec{q}
\nn \\
-it_{\Lambda \Sigma^{*} \pi}= \frac{6}{5}\frac{D+F}{2f}\vec{S} \cdot \vec{q}
\ea
\end{center}
\noindent for an incoming pion of momentum $\vec{q}$.

The calculation of the weak $\Sigma N \pi$ and $\Sigma^{*} N \pi$ vertices is not
so trivial. Here we only care about the parity conserving part of these
vertices,
as said before. As commented in the former section one cannot use current
algebra arguments and must resort to some model. Here we take a simple one
using $SU(3)$ symmetry plus the $\Delta
I=1/2$ rule. We implement this rule by introducing in the initial state a
fictitious $|1/2 \  1/2;
1\rangle \equiv \bar{K}^{0}$ state, which we couple to the nucleon, and then we
couple the resulting state to the pion to get the $\Sigma$ ($\Sigma^{*}$)
baryon. Experimental values for the $\Sigma N \pi$ couplings (both for the
parity conserving and parity violating parts) and some
constraints can be found in
reference~\cite{Bando:1988pn}. Our results satisfy the aforementioned constraints since they come from
$SU(2)$ symmetry plus $\Delta I=1/2$ rule, and therefore are also valid for the
$\Sigma^{*} N \pi$ couplings. We get also a reasonable good agreement with the
experimental data and this is enough because, as we are going to see, there are
big cancellations between the diagrams with an intermediate $\Sigma$ and those
with an intermediate $\Sigma^{*}$, and the final results barely depend on the
inclusion of these diagrams in the calculations. In fact, in our final results we
will neglect these diagrams.

We get for the weak vertex of the diagrams of fig.~\ref{vertex}:

\begin{center}
\be
\label{vertexsigweak}
-it_{\Sigma N \pi}=\mathfrak{R} (a_{\Sigma}d+b_{\Sigma}f) \vec{\sigma} \cdot \vec{q}
\hspace{1.6cm}
-it_{\Sigma^{*} N \pi}=\mathfrak{R} a_{\Sigma^{*}}d \vec{S}^{\dag} \cdot \vec{q} 
\ee
\end{center}
\noindent where the values of $a_{\Sigma}$, $b_{\Sigma}$ and $a_{\Sigma^{*}}$
are given in table~\ref{sigco}.

\begin{table}[ht]
\begin{center}
\begin{tabular}{|c|ccccc|}
\hline
 \ \ & $\pi^{-} p$ & $\pi^{0} p$ & $\pi^{-} n$ & $\pi^{0} n$
 & $\pi^{+} n$\\ 
\hline
 $a_{\Sigma}$ & 0 & 0 & \scriptsize{-1/$\sqrt{10}$} &
 \scriptsize{-1/$\sqrt{10}$} & \scriptsize{-1/$\sqrt{10}$} \\
 $b_{\Sigma}$ & \scriptsize{-1/$\sqrt{3}$} & \scriptsize{1/$\sqrt{3}$} &
 \scriptsize{-1/$\sqrt{6}$} & 0 & \scriptsize{1/$\sqrt{6}$} \\
 $a_{\Sigma^{*}}$ & $\frac{3\sqrt{2}}{5\sqrt{15}}$ &
 $-\frac{3\sqrt{2}}{5\sqrt{15}}$ & $\frac{3}{5\sqrt{15}}+\frac{3}{5\sqrt{5}}$ &
 $\frac{3}{5\sqrt{5}}$ & $-\frac{3}{5\sqrt{15}}+\frac{3}{5\sqrt{5}}$\\
 \hline 
\end{tabular}
\caption{Values of the coefficients of eq.~(\ref{vertexsigweak}) for the parity
conserving
part of the weak vertex.}
\end{center}
\label{sigco} 
\end{table}

With these values for the $\Sigma N \pi$ and $\Sigma^{*} N \pi$ couplings we can
finally calculate the $v_{\Sigma}(q)$ and
$v_{\Sigma^{*}}(q)$ functions. We obtain:

\begin{center}
\ba
\label{vsigma}
v_{\Sigma}(q)=\int
\frac{d^{3}p}{(2\pi)^{3}}\frac{D}{\sqrt{3}f}
\left(\frac{D+F}{2f}\right)(\vec{p}^{\ 2}+\vec{p}\cdot
\vec{q})\frac{M_{\Sigma}}{2E_{\Sigma}}\frac{1}{\omega
\omega'}\ \frac{1}{\omega+\omega'}\nn \\ \times \frac{1}
{E_{\Sigma}+\omega-M_{\Lambda}}\ \frac{1}{E_{\Sigma}+\omega'-M_{\Lambda}}
(\omega+\omega'+E_{\Sigma}-M_{\Lambda}) \nn \\
v_{\Sigma^{*}}(q)=-\int \frac{d^{3}p}{(2\pi)^{3}}
\frac{12\sqrt{2}}{25}\left(\frac{D+F}{2f}\right)^{2}(\vec{p}^{\ 2}+\vec{p}\cdot
\vec{q})\frac{M_{\Sigma^*}}{2E_{\Sigma^*}}\frac{1}{\omega
\omega'}\ \frac{1}{\omega+\omega'}\nn \\ \times \frac{1}
{E_{\Sigma^{*}}+\omega-M_{\Lambda}}\ \frac{1}{E_{\Sigma^{*}}+\omega'-M_{\Lambda}}
(\omega+\omega'+E_{\Sigma^{*}}-M_{\Lambda})
\ea
\end{center}

\noindent where the notation followed is the same as in eq~(\ref{vn}). We can
see that $v_{\Sigma}(q)$ and $v_{\Sigma^{*}}(q)$ have opposite
sign, thus leading to the aforementioned cancellation.

At this point we have all the ingredients to calculate the $V^{2\pi}_{corr}(q)$
of equation
eq~(\ref{tt}) corresponding to the correlated two-pion exchange. The only thing
one has to do is to generalize the definition of $v(q)$ of
eq.~(\ref{rv}) by:

\begin{center}
\be
\label{vtot}
v(q)=v_{N}(q)+v_{\Delta}(q)+v_{\Sigma}(q)+v_{\Sigma^{*}}(q)
\ee
\end{center}

\begin{figure}[ht]
\centerline{\includegraphics[width=0.8\textwidth]{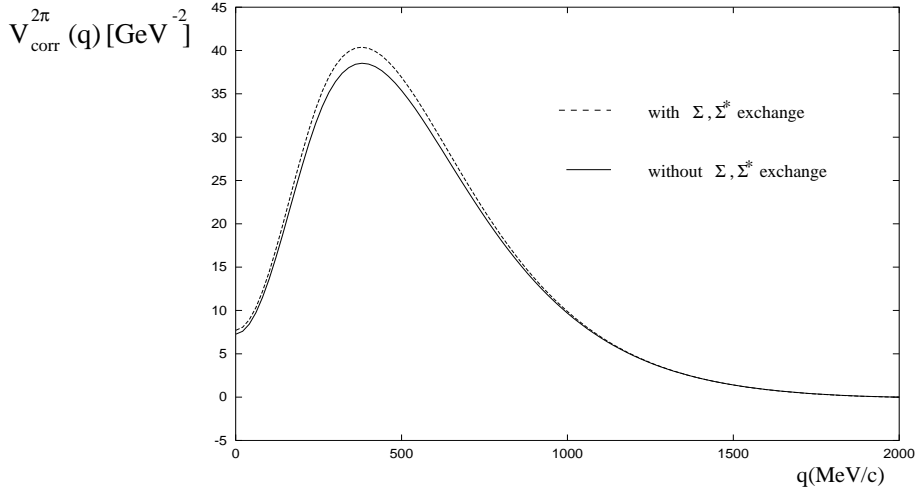}}
\caption{The $V^{2\pi}_{corr}(q)$ function corresponding to the correlated two-pion exchange with
and without including the $\Sigma$, $\Sigma^{*}$ exchange diagrams, before
multiplying by the $\mathfrak{R}$ factor.}
\label{t}
\end{figure}

In fig.~\ref{t} we plot $V^{2\pi}_{corr}(q)$ both with and without considering the
diagrams with intermediate $\Sigma$, $\Sigma^{*}$. As we can see, the effect of
these latter diagrams is very small due to the cancellation between them. 
 In our calculations we will no
 longer include these diagrams.
 
In order to include the short range correlations we will use 
$V^{2\pi}_{corr}(q)-\widetilde{V}^{2\pi}_{corr}(q)$
instead of $V^{2\pi}_{corr}(q)$, where the tilde on the function means that $\vec{q}^{\textrm{
}2}$ must
be replaced by $\vec{q}^{\textrm{ }2}+q_{c}^{2}$, being $q_c$ the inverse of a typical
correlation length (see Appendix).

\subsection{Uncorrelated two-pion exchange}
\begin{figure}[ht]
\centerline{\includegraphics[width=0.5\textwidth]{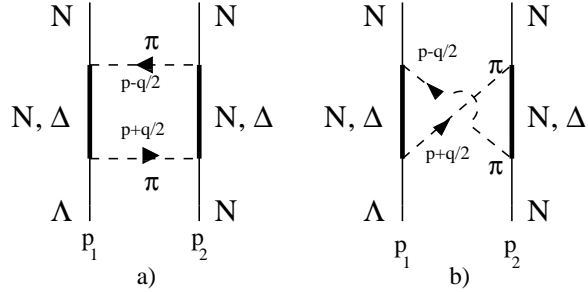}}
\caption{The two pion uncorrelated exchange triangle vertex through $N$ and
$\Delta$ excitations.}
\label{uncorr}
\end{figure}

The other set of processes that we have to study when considering the two-pion
exchange is the one in which the exchanged pions only interact with baryonic
legs and not with other pions (uncorrelated exchange). The corresponding Feynman
diagrams are depicted in fig.~\ref{uncorr}.

We do not include the diagrams with an intermediate $\Sigma$  and
$\Sigma^{*}$,
 because one expects a similar cancellation to the one found in the correlated
 exchange, nor the diagrams with two nucleon propagators in diagram 
 \ref{uncorr} (a), which
 correspond to final state interaction and are included in the correlations. 
  We will neglect also the spin dependent term, which is found to be of
 order $2q^{2}/(9p^{2})$ with respect to the spin-independent one. Here $q$ is
 the momentum transfer and $p$ is the loop variable, which is cut around 1GeV by
 the cut off, so taking $\langle p\rangle \sim 800$ MeV/c and $q\sim 400$
 MeV/c we find $2q^{2}/(9p^{2})\sim 6\%$, and therefore we do not care about it.

Let us consider first the direct exchange. As an example we write the
contribution of the diagram with one intermediate nucleon in the left hand 
 baryonic line
of fig.~\ref{uncorr} (a), and an intermediate $\Delta$ in the right one. We will
include at the end the factor $\mathfrak{R}$ associated to the weak vertex. The
modifications needed to describe the other diagrams are trivial. The potential
associated to the aforementioned exchange is:
\small{
\begin{center}
\ba
\label{dirbox}
\lefteqn{-iv^{dir}_{N\Delta}(q)=\int
\frac{d^{4}p}{(2\pi)^{4}}\frac{M_{N}}{E(\vec{p_{1}}-\vec{p}-\vec{q}/2)}\ \frac{M_{\Delta}
}{E_{\Delta}(\vec{p_{2}}+\vec{p}+\vec{q}/2)}\ \frac{i\textrm{ }(1-n(\vec{p_{1}}
-\vec{p}-\vec{q}/2))}{p_{1}^{0}-p^{0}-q^{0}/2-E(\vec{p_{1}}
-\vec{p}-\vec{q}/2)+i\epsilon}}\nn \\ 
& & {}\times
\frac{i}{p_{2}^{0}+p^{0}+q^{0}/2-E_{\Delta}(\vec{p_{2}}+\vec{p}+\vec{q}/2)+i\epsilon}\ \frac{i}
{(p+q/2)^{2}-\mu^{2}+i\epsilon}\ \frac{i}{(p-q/2)^{2}-\mu^{2}+i\epsilon}
\left(\frac{f_{\pi NN}}{\mu}\right)^{2}\nn\\ & & {}\times\ \left(\frac{f_{\pi \Delta N}}
{\mu}\right)^{2}\vec{\sigma_{1}}\cdot(\vec{p}-\vec{q}/2)\ \vec{\sigma_{1}}
\cdot(\vec{p}+\vec{q}/2)\ \vec{S_{2}}\cdot(\vec{p}-\vec{q}/2)\ \vec{S_{2}}^{\dagger}\cdot(\vec{p}+\vec{q}/2) \tau_{1}^{i}
\tau_{1}^{j} T_{2}^{i} T_{2}^{\dagger j}
\ea
\end{center}}
\normalsize{

\noindent where} $\vec{S}$ and $\vec{T}$ are the spin and isospin transition
operators normalized such that they satisfy:

\begin{center}
\ba
\label{SyT}
S_{i}S_{j}^{\dagger}=\frac{2}{3}\delta_{ij}-\frac{i}{3}\epsilon_{ijk}
\sigma_{k} \nn \\
T_{i}T_{j}^{\dagger}=\frac{2}{3}\delta_{ij}-\frac{i}{3}\epsilon_{ijk}
\tau_{k}
\ea
\end{center}

To regularize these loops we will include a cutoff in the space of intermediate
states together with static form factors in the $\pi NN$ vertices, as done in
the case of correlated two pion exchange. We will also work with the initial $\Lambda$
and nucleon at rest in order to simplify the calculations.

The calculation of the crossed exchange is analogous. As before, we will write
the contribution of the diagram containing one nucleon propagator in the left
baryonic line of fig.\ref{uncorr}.b, and a $\Delta$ propagator in the right
one:

\small{
\begin{center}
\ba
\label{crossbox}
\lefteqn{-iv^{cross}_{N\Delta}(q)=\int
\frac{d^{4}p}{(2\pi)^{4}}\frac{M_{N}}{E(\vec{p_{1}}-\vec{p}-\vec{q}/2)}\ \frac{M_{\Delta}
}{E_{\Delta}(\vec{p_{2}}-\vec{p}+\vec{q}/2)}\ \frac{i\textrm{ }(1-n(\vec{p_{1}}
-\vec{p}-\vec{q}/2))}{p_{1}^{0}-p^{0}-q^{0}/2-E(\vec{p_{1}}
-\vec{p}-\vec{q}/2)+i\epsilon}}\nn \\ 
& & {}\times
\frac{i}{p_{2}^{0}-p^{0}+q^{0}/2-E_{\Delta}(\vec{p_{2}}-\vec{p}+\vec{q}/2)+i\epsilon}\ \frac{i}
{(p+q/2)^{2}-\mu^{2}+i\epsilon}\ \frac{i}{(p-q/2)^{2}-\mu^{2}+i\epsilon}
\left(\frac{f_{\pi NN}}{\mu}\right)^{2}\nn\\& &{}\times\left(\frac{f_{\pi \Delta
N}}{\mu}\right)^{2}\vec{\sigma_{1}}\cdot(\vec{p}-\vec{q}/2)\ \vec{\sigma_{1}}
\cdot(\vec{p}+\vec{q}/2)\ \vec{S_{2}}\cdot(\vec{p}-\vec{q}/2)\ \vec{S_{2}}^{\dagger}\cdot(\vec{p}+\vec{q}/2) \tau_{1}^{i}
\tau_{1}^{j} T_{2}^{i} T_{2}^{\dagger j}
\ea
\end{center}}
\normalsize{

The} regularization is achieved also by including a cutoff and static form
factors. The $p^{0}$ integral can be analytically
performed in both cases.

The resulting uncorrelated two-pion exchange potential can be divided into an
isoscalar and an isovector piece:

\begin{center}
\be
\label{descomp}
V^{2\pi}_{unc}(q)=V^{2\pi}_{is,unc}(q)+V^{2\pi}_{iv,unc}(q)\vec{\tau}\cdot\vec{\tau}
\ee
\end{center}

\noindent where these two pieces are:

\begin{center}
\ba
\label{totcross}
V^{2\pi}_{is,unc}(q) =
3\mathfrak{R}v_{NN}^{cross}(q)+\frac{4\mathfrak{R}}{3}\left(v^{dir}_{\Delta
N}(q)+v^{dir}_{N\Delta}(q)+v^{cross}_{\Delta N}(q)+v^{cross}_{N\Delta}(q)\right)+
\nonumber \\ \frac{16\mathfrak{R}}{27}\left(v^{dir}_{\Delta\Delta}(q)+v^{cross}_{\Delta\Delta}(q)\right)\nn \\
V^{2\pi}_{iv}(q) = 2\mathfrak{R}v_{NN}^{cross}(q)+\frac{4\mathfrak{R}}{9}\left(v^{dir}_{\Delta
N}(q)+v^{dir}_{N\Delta}(q)-v^{cross}_{\Delta N}(q)-v^{cross}_{N\Delta}(q)\right) +
\nonumber \\ \frac{8\mathfrak{R}}{81}\left(v^{cross}_{\Delta\Delta}(q)-v^{dir}_{\Delta\Delta}(q)\right)
\ea
\end{center}

\begin{figure}[ht]
\centerline{\includegraphics[width=0.8\textwidth]{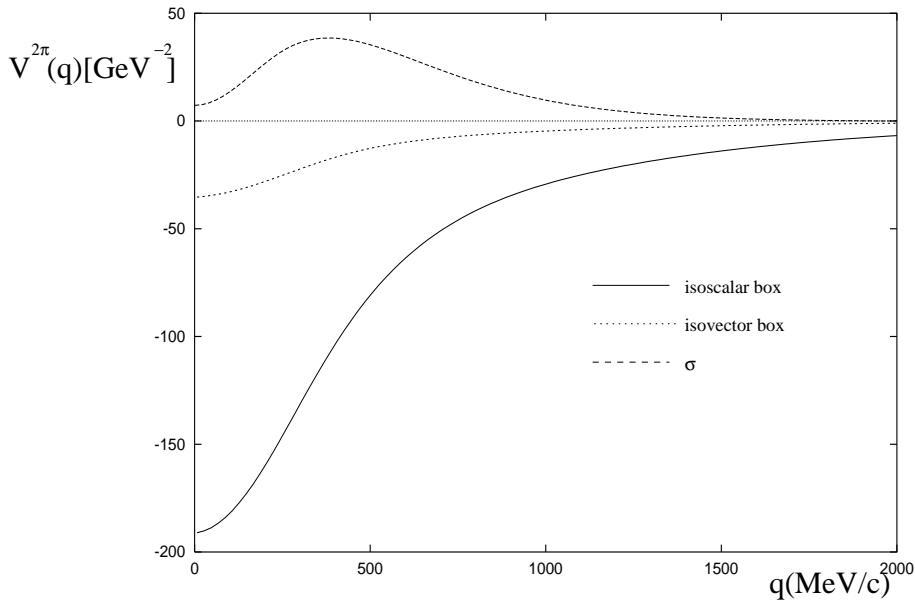}}
\caption{Isoscalar and isovector pieces of the uncorrelated two-pion exchange
potential and the $\sigma$ (correlated $2\pi$ exchange) potential (without including the factor
$\mathfrak{R}$).}
\label{boxplot}
\end{figure}

It is worth mentioning that, as we are not taking into account direct diagrams with two
nucleon propagators, $V^{2\pi}_{unc}(q)$ is real. The isoscalar and isovector parts
of the function of eq.~(\ref{totcross}) are plotted in fig.~\ref{boxplot}. As we can see there,
the isoscalar part is dominant around $q\sim 420$ MeV/c, so from now on we will neglect the isovector
contribution. This simplifies the description since the remaining
scalar-isoscalar contribution can be added directly to the correlated
"$2\pi/\sigma$" exchange of the previous subsection. As in the correlated
exchange case, short range correlations must be taken into account and their
inclusion is achieved, as before, by subtracting to $V^{2\pi}_{is,unc}(q)$ its
value at $\vec{q}^{\ 2}+q_{c}^{2}$ (see Appendix).% In fig. \ref{twopion} we
%plot the contributions of correlated and uncorrelated two-pion exchange, and the
%sum of both contributions. In the sum we have subtracted its value at
%$q_{c}=780$ MeV.

Up to now we have not taken into account the parity violating part of the weak
vertex in the two-pion exchange contributions. We have evaluated this contribution both in the correlated and
uncorrelated two-pion exchange and we find these terms small. Additionally 
there is a cancellation between them, so the
inclusion of this term is irrelevant. The point is that one cannot have an
intermediate $\Delta$ coupled to the $\Lambda$ in the weak vertex, and this reduces
the correlated exchange contribution to $1/5$ of the one of the parity
conserving part but also reduces strongly the contribution of the uncorrelated
exchange, leading to a very big cancellation at the relevant momentum 
 $q\sim 420$ MeV/c. The structure of these terms is exactly the same as the
 (s-wave) parity violating term in the $K$ exchange and their summed strength is
 found negligible compared to this latter term.

 The inclusion of the two-pion scalar-isoscalar terms is easy. These terms only have
 interferences with the p-wave part of the $\pi +K$ exchange. By using
 eqs. (\ref{tt}) and (\ref{totcross}) we define:
 
 \begin{center}
 \begin{equation}
 V'_{2\pi}(q)=\frac{V^{2\pi}_{corr}(q)+V^{2\pi}_{is ,unc}(q)}{\mathfrak{R}}
\end{equation}
\end{center}

\noindent and the new contributions are accounted for by adding to
eqs. (\ref{38}) 
to (\ref{41}) the following terms:

\begin{eqnarray}
\lefteqn{T^{2\pi}_{p,S}(q) = (V'_{2\pi}(q))^{2}} \nonumber \\
\lefteqn{T^{2\pi}_{p,P,int}(q) = -V'_{2\pi}(q)(2V'_{l}+V'_{l,K} \Ckmpp)
\textrm{Re}\left(\frac{1}{1-UV_{l}}\right) {} }  \nonumber \\
 & & {} -V'_{2\pi}(q)(4V'_{t}+2V'_{t,K} \Ckmpp)
\textrm{Re}\left(\frac{1}{1-UV_{t}}\right) \nonumber \\
\lefteqn{T^{2\pi}_{n,S}(q) = \frac{1}{2}(V'_{2\pi}(q))^{2}} \nonumber \\
\lefteqn{T^{2\pi}_{n,P,int}(q) = -V'_{2\pi}(q)(V'_{l}+V'_{l,K} \Ckznp)
\textrm{Re}\left(\frac{1}{1-UV_{l}}\right) {} }  \nonumber \\
 & & {} -V'_{2\pi}(q)(2V'_{t}+2V'_{t,K} \Ckznp)
\textrm{Re}\left(\frac{1}{1-UV_{t}}\right) 
\end{eqnarray}

% For revised version

\begin{figure}[h]
\centerline{\includegraphics[width=0.7\textwidth]{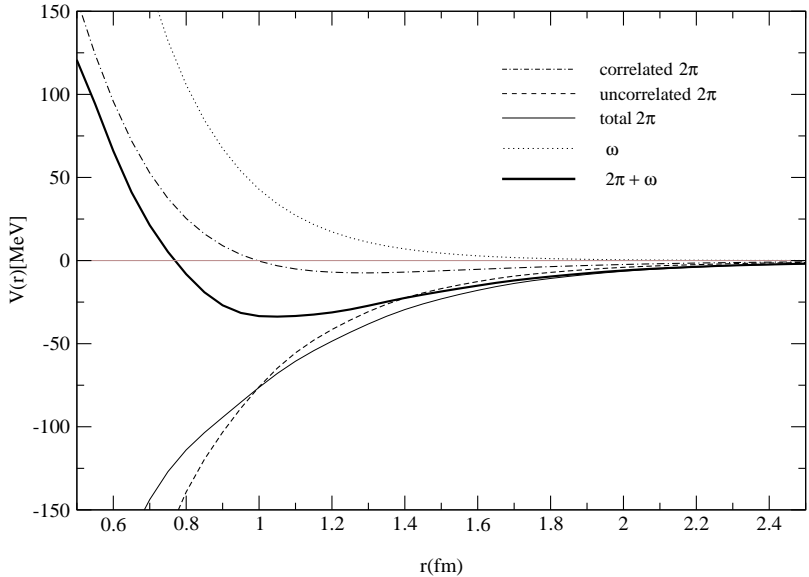}}
\caption{Central potentials of the strong $NN$ interaction in coordinate
space from correlated two-pion (dashed-dotted line), 
uncorrelated two-pion (dashed line) and omega meson (dotted line)
exchanges. The thick solid and thin solid
lines denote the sum of all contributions
and the sum without the omega meson, respectively.}
\label{fig:NNpotcor}
\end{figure}

The repulsive character of our correlated two-pion exchange in momentum
space is somewhat unexpected but it is linked to an important
constraint of chiral symmetry, i.e., the Adler zero at $s=\mu^2 /2$
where the $\pi\pi$ scalar-isoscalar amplitude changes sign. In
coordinate space it leads to a moderate attraction at intermediate and
long distances, and to a repulsion at short
distances~\cite{Oset:2000gn}. Such repulsion from the correlated
two-pion exchange is actually not so novel and in models like the
Skyrme model, where only pion degrees of freedom are considered, the
scalar-isoscalar $NN$ interaction also shows repulsion at short
distances~\cite{Jackson:1985bn}. However, it is worth noting that when
we sum the correlated and uncorrelated two-pion contribution we obtain
a net attraction in coordinate space. This is shown in fig. \ref{fig:NNpotcor}.

Once at this point, we would also like to connect our results with 
present realistic forces of the $NN$ interactions. We choose the Argonne
potentials $v_{14}$, $v_{18}$~\cite{Argonne} (see also the
paper \cite{Robilotta:1997ji} for a discussion of this central
potential). We observe that this potential has a moderate attraction
of around 20-30 MeV at intermediate distances and becomes repulsive at
short distances. Our potential with correlated plus uncorrelated
two-pion exchange is only attractive. Obviously other contributions
are missing there, and the conventional meson exchange approach to
generate this repulsion is the $\omega$ exchange. We thus add the omega
exchange to our potential and find that for standard values of the
coupling and the form factor we can reproduce fairly well the Argonne
$v_{14}$ potential. This is seen in fig.~\ref{fig:NNpotcor}, where
we have chosen the coupling $g_\omega = 13$ and $\Lambda_\omega=1.4$ GeV, for
a monopole form factor.

Now we turn to the weak interaction. Since we have seen that $\omega$
exchange plays a role in the strong interaction, we would also like to
include it in the weak transition. For this purpose we take the strong
coupling determined here from the $NN$ central potential and
use the weak coupling of the $\omega$ given in \cite{PRB97}
$g^W_{\Lambda N \omega} = 3.69 G\mu^2$ (for the parity conserving
part, and the sign is
the same one as for the weak parity conserving pion coupling), and same
form factor as for the strong vertex. With the sign prescription for our
strong and weak Lagrangians this $\omega$ potential in momentum space
is given by
\begin{equation}
   V_\omega(\vec{q}) = {g^W_{\Lambda N \omega} g_\omega \over
   \vec{q}^{\ 2} + m_\omega^{2}} \left( {\Lambda_\omega^2 - m_\omega^2
   \over \Lambda_\omega^2 + \vec{q}^{\ 2}} \right)^2
\end{equation}

\begin{figure}[bh]
\centerline{\includegraphics[width=0.7\textwidth]{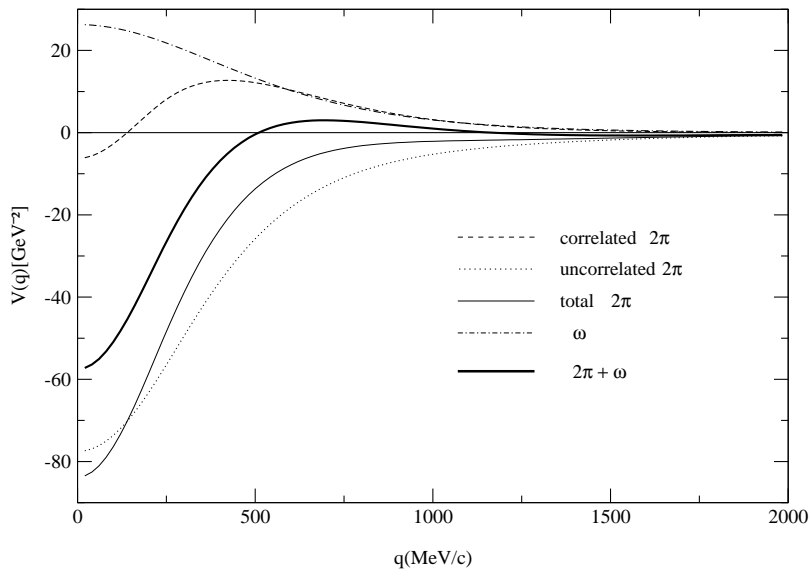}}
\caption{Weak transition potentials 
in momentum space (divided by $G\mu^{2}$) from correlated two-pion (dashed line), 
uncorrelated two-pion (dotted line) and omega meson (dotted-dashed line)
exchanges. The thick solid and thin solid lines 
denote the sum of all contributions
and the sum without the omega meson, respectively. Here we have multiplied by
the factor $\mathfrak{R}$ both the correlated and uncorrelated two-pion
exchange, and we have already included the effect of 
short range correlations by subtracting
to the potentials their values at $\vec{q}^{\ 2}+q_{c}^{2}$.  }
\label{fig:omegaplot}
\end{figure}

In fig.~\ref{fig:omegaplot} we show now our results for the transition
potential in momentum space for the weak two-pion exchange, omega
exchange and their sum. We observe that around $q=400 {\rm MeV/c}$ the
inclusion of the $\omega$ exchange reduces the contribution of the
scalar-isoscalar transition to about 30\% of the contribution of the
two-pion exchange alone. We shall discuss the effect of the $\omega$
exchange in the results for the total decay widths and
$\Gamma_{n}/\Gamma_{p}$ ratio in the next section.

%%%%%%%%%%%%%%%%%%%%%%%%%%%%%%%%%%%%%%%%%%%%%%%%%

%\begin{figure}[ht]
%\centerline{\includegraphics[width=0.92\textwidth]{sigboxfig.ps}}
%\label{twopion}
%\end{figure}
\section{Results}

We show the results for the proton and neutron induced decay widths of $^{12}_{\Lambda}$C separating the different
contributions. In table~\ref{ope} we show the results obtained with only one
pion exchange. In table~\ref{oke} we show the results with only one kaon
exchange and only the scalar-isoscalar two-pion exchange. In table~\ref{interf} we
show the different interference terms and in table~\ref{tot} the results obtained
including $\pi+K$, $\pi+K+2\pi$ and $\pi +K+ 2\pi+ \omega$.

\begin{table}[ht]
\begin{center}
\begin{tabular}{|c|ccccc|}
\hline 
 \ \ & $S$ & $P_{L}$ & $P_{T}$ & $P_{int.\ LT}$
 & Total\\ 
\hline
 $\Gamma_{p}/\Gamma^{free}_{\Lambda}$ & 0.177 & 0.684 & 0.012 & 0.082 & 0.956 \\
 $\Gamma_{n}/\Gamma^{free}_{\Lambda}$ & 0.089 & 0.049 & 0.003 & -0.021 & 0.119 \\
\hline
\end{tabular}
\caption{OPE contributions to $\Gamma_{p}$ and $\Gamma_{n}$: s-wave,
longitudinal p-wave, transverse p-wave
and interference between longitudinal and transverse p-wave. Decay rates given in
units of the free $\Lambda$ width.}
\label{ope}
\end{center}
\end{table}

\begin{table}[ht]
\begin{center}
\begin{tabular}{|c|cccccc|}
\hline 
 \ \ & $S$ & $P_{L}$ & $P_{T}$ & $P_{int.\ LT}$
 & Total kaon& $2\pi$\\ 
\hline
 $\Gamma_{p}/\Gamma^{free}_{\Lambda}$ & 0.058  & 0.152 & 0.012 & 0.030 & \ \ 0.253 & 0.191\\
 $\Gamma_{n}/\Gamma^{free}_{\Lambda}$ & 0.110 & 0.022 & 0.005 & -0.019 & \ \ 0.118 & 0.095\\
\hline
\end{tabular}
\caption{One kaon exchange and two pion exchange contributions to the partial
decay widths.}
\label{oke}
\end{center}
\end{table}

\begin{table}[ht]
\begin{center}
\begin{tabular}{|c|cccccccc|}
\hline 
 \ \ & $\pi K_{S}$ & $\pi K_{L}$ & $\pi K_{T}$ & $\pi K_{LT}$
 & $"\sigma"\pi_{L}$ & $"\sigma"\pi_{T}$ & $"\sigma"K_{L}$ & $"\sigma"K_{T}$\\ 
\hline
 $\Gamma_{p}/\Gamma^{free}_{\Lambda}$ & 0.075  & -0.629 & -0.024 & -0.109 & -0.264 &
 0.057 &
 0.126 & -0.060 \\
 $\Gamma_{n}/\Gamma^{free}_{\Lambda}$ & 0.065 & -0.064 & -0.007 & 0.042 & -0.132 &
 0.029 & 0.082
 & -0.039\\
\hline
\end{tabular}
\caption{Interferences between the different contributions.
Here $"\sigma"$ means two pion exchange, both correlated and uncorrelated.}
\label{interf}
\end{center}
\end{table}

\begin{table}[ht]
\begin{center}
\begin{tabular}{|c|cccc|}
\hline 
 \ \ & $\pi$ & $\pi+K$ & $\pi+K+"\sigma"$ & $\pi+K+"\sigma"+\omega$\\ 
\hline
 $\Gamma_{p}/\Gamma^{free}_{\Lambda}$ & 0.956 & 0.522 & \ \ \ 0.571& 0.504\\
 $\Gamma_{n}/\Gamma^{free}_{\Lambda}$ & 0.119 & 0.273 & \ \ \ 0.308& 0.265\\
\hline
\end{tabular}
\caption{Decay rates obtained when considering $\pi$, $\pi +K$, $\pi
+K$+two-pion and $\pi + K$+two-pion+$\omega$ 
exchange.}
\label{tot}
\end{center}
\end{table}

% Revised version: with Omega exchange
\begin{table}[ht]
\begin{center}
\begin{tabular}{|cccccc|}
\hline 
 Nucleus & $\Gamma_p/\Gamma^{free}_{\Lambda}$ &
 $\Gamma_n/\Gamma^{free}_{\Lambda}$ & $(\Gamma_{p}+
 \Gamma_{n})/\Gamma^{free}_{\Lambda}$ & $\Gamma_{n}/\Gamma_{p}$ &
 $\Gamma_{tot}/\Gamma^{free}_{\Lambda}$\\ 
\hline
 $^{12}_{\Lambda}$C   & 0.504 & 0.265 & 0.769 & 0.53 & 1.289\\
 $^{28}_{\Lambda}$Si  & 0.665 & 0.351 & 1.016 & 0.53 & 1.386\\
 $^{40}_{\Lambda}$Ca  & 0.694 & 0.366 & 1.060 & 0.53 & 1.390\\
 $^{56}_{\Lambda}$Fe  & 0.754 & 0.398 & 1.152 & 0.53 & 1.452\\
 $^{89}_{\Lambda}$Y   & 0.785 & 0.414 & 1.199 & 0.53 & 1.499\\
 $^{139}_{\Lambda}$La & 0.765 & 0.403 & 1.168 & 0.53 & 1.468\\
 $^{208}_{\Lambda}$Pb & 0.847 & 0.446 & 1.293 & 0.53 & 1.593\\
\hline
\end{tabular}
\caption{Decay rates and the $\Gamma_{n}/\Gamma_{p}$ ratio for different
hypernuclei. In $\Gamma_{tot}$ we have included the contributions from the
mesonic decay and the $2p2h$ channel.}
\label{final}
\end{center}
\end{table}

\begin{table}[ht]
\begin{center}
\begin{tabular}{|ccc|}
\hline 
 Nucleus & $\Gamma/\Gamma^{free}_{\Lambda}$ & Experiment \\ 
\hline
 & &  \\
% \ \ $^{3}_{\Lambda}$H & $3^{+7}_{-1.3}$\cite{6s} & Emulsion \\
%\vspace{0.08cm} 
%   & $0.9^{+0.4}_{-0.3}$\cite{7s} & Emulsion \\
%\vspace{0.08cm} 
%   & $2.1^{+0.6}_{-0.4}$\cite{8s} & Emulsion \\
%\vspace{0.08cm} 
%   & $1.0^{+0.3}_{-0.2}$\cite{9s} & Bubble chamber \\
%\vspace{0.08cm} 
%   & $1.1^{+0.3}_{-0.2}$\cite{10s} & Bubble chamber \\
%\vspace{0.08cm} 
% \ \ $^{4}_{\Lambda}$H & $1.5^{+2}_{-0.6}$\cite{6s} &  Emulsion \\
%\vspace{0.08cm} 
%   & $0.7^{+1}_{-0.3}$\cite{11s} & Emulsion \\
%\vspace{0.08cm} 
%   & $1.0^{+0.6}_{-0.4}$\cite{7s} & Emulsion \\
% \vspace{0.08cm} 
%  & $1.36^{+0.18}_{-0.19}$\cite{12s,13s} & $(K^{-}_{stop},\pi^{-})$ \\
%\vspace{0.08cm} 
% \ \ $^{4}_{\Lambda}$He & $1.2^{+1.2}_{-0.7}$\cite{7s} &  Emulsion \\
%\vspace{0.08cm} 
%   & $1.07\pm 0.11$\cite{14s} & $(K^{-},\pi^{-})$ \\
%\vspace{0.08cm} 
%   & $1.03\pm 0.12$\cite{13s} & $(K^{-}_{stop},\pi^{-})$ \\
%\vspace{0.08cm} 
%  \ \ $^{5}_{\Lambda}$He & $1.9^{+2.7}_{-0.7}$\cite{6s} &  Emulsion \\
%\vspace{0.08cm} 
%   & $1.5^{+1.3}_{-0.5}$\cite{11s} & Emulsion \\
% \vspace{0.08cm} 
%  & $1.0^{+0.8}_{-0.3}$\cite{7s} & Emulsion \\
%\vspace{0.08cm} 
%   & $0.96^{+0.22}_{-0.18}$\cite{15s} & Emulsion \\
%\vspace{0.08cm} Bhang:1998zp
%   & $1.03\pm 0.09$\cite{Szymanski:1991ik}& $(K^{-},\pi^{-})$ \\
  
\ \ $^{9}_{\Lambda}$Be & $1.31\pm 0.21$\cite{Szymanski:1991ik} &  $(K^{-},\pi^{-})$ \\ 
   $^{11}_{\Lambda}$B & $1.37\pm 0.17$\cite{Szymanski:1991ik,Grace:1985yi} & $(K^{-},\pi^{-})$ \\
   & $1.25\pm 0.08$\cite{Park:2000zp} & $(\pi^{+},K^{+})$ \\
   $^{12}_{\Lambda}$C & $1.25\pm 0.19$\cite{Szymanski:1991ik,Grace:1985yi} & $(K^{-},\pi^{-})$ \\
   & $1.14\pm 0.08$\cite{Bhang:1998zp} & $(\pi^{+},K^{+})$ \\
   $^{16}_{\Lambda}$Z & $3.1^{+1.2}_{-0.9}$\cite{Nield:1976un} & $^{16}$O beam, $K^{+}$
   tagging \\
   $^{27}_{\Lambda}$Al & $1.30\pm 0.07$\cite{Park:2000zp} & $(\pi^{+},K^{+})$ \\
   $^{28}_{\Lambda}$Si & $1.28\pm 0.08$\cite{Park:2000zp} & $(\pi^{+},K^{+})$ \\
   $_{\Lambda}$Fe & $1.22\pm 0.08$\cite{Park:2000zp} & $(\pi^{+},K^{+})$ \\
   $\bar{p}+^{209}$Bi & $1.1^{+1.1}_{-0.4}$\cite{Bocquet:1987kw} & Delayed fission \\
    & $1.5\pm 0.3\pm 0.5$\cite{Armstrong:1993xi} & Delayed fission \\
   $p+^{209}$Bi & $1.8\pm 0.1\pm 0.3$\cite{Kulessa:1998kg} & Delayed fission \\
   $\bar{p}+^{238}$U & $2.6^{+2.2}_{-1.1}$\cite{Bocquet:1987kw} & Delayed fission \\
    & $2.0\pm 0.5\pm 0.5$\cite{Armstrong:1993xi} & Delayed fission \\
   $p+^{238}$U  & $1.1\pm 0.3$\cite{Ohm:1997td}   & Delayed fission \\ 
                & $1.4\pm 0.4$\cite{Zychor} & re-analysis of  \cite{Ohm:1997td}\\
   $p+$(Au, U, Bi) & $2.0\pm 0.2\pm 0.2$\cite{Kamys:2000gt} & Delayed fission \\
   
\hline
\end{tabular}
\caption{Experimental values of the total width for different nuclei. The value
for $_{\Lambda}$Fe represents for the average width of $^{55}_{\Lambda}$Mn, 
$^{55}_{\Lambda}$Fe and $^{56}_{\Lambda}$Fe.}
\label{park}
\end{center}
\end{table}
\normalsize

While $K$ or $2\pi$ contributions by themselves are small compared to
$\Gamma_{p}$ from OPE, the interference effects with the OPE contribution are
large. As we can see from table~\ref{interf}, one of the important pieces of
interference is in the longitudinal part of the p-wave contribution. The kaon
exchange produces a large cancellation of this contribution from OPE in
$\Gamma_{p}$. The interference of $2\pi$ exchange and OPE in the p-wave
longitudinal channel is also relevant, and to a smaller extend also the
interference of $2\pi$ with $K$ exchange.

If one looks at table~\ref{interf} and compares it to table~\ref{ope} one can see
that the main effect of the $K$ exchange contribution is to decrease the p-wave
contribution of the OPE in the proton case, which was responsible for the large
$\Gamma_{n}/\Gamma_{p}$ ratio. The two-pion exchange contribution has finally a
small contribution to the rates because of the cancellation between direct
$2\pi$ and interference contributions.

It is worth recalling, as we saw from fig.~\ref{boxplot}, that the $\sigma$ and
uncorrelated $2\pi$ contributions have different signs and there are large
cancellations between them at the relevant momentum $q\sim 420$ MeV/c. Let us
stress once more that we obtain a sign for the $\sigma$ exchange here which is
opposite to the conventional one.
Should we have the $\sigma$ contribution with opposite sign to ours and about
the same strength, the combination of $\sigma$ and uncorrelated two-pion
exchange would give a contribution for the $2\pi$ part alone about 6 times bigger
than here, and this would render the total rates unacceptably large in spite of
the interference terms, which are only multiplied by a factor 2.5.

%%% for revised version
We address now the contribution from the $\omega$ exchange. We have
already seen in table~\ref{tot} that the combined effect of the
two-pion exchange (``$\sigma$'' in the table) is very moderate. It
increases $\Gamma_p$ by 10\% and $\Gamma_n$ by 13 \%. The inclusion of
$\omega$, as looks clear from fig.~\ref{fig:omegaplot}, should give
contributions of the same order of magnitude. This is the case indeed,
and in table~\ref{tot} we show the contribution when the $\omega$
exchange is added, which indeed is small but helps one to obtain a
better agreement with experiment for the total widths.
%%%%

In table~\ref{final} we present the results for $\Gamma_p$, $\Gamma_n$ and the
$\Gamma_{n}/\Gamma_{p}$ ratio for different nuclei. We have used the parameters
of ref.~\cite{Dubach:1996dg} for the $NNK$ coupling. We find that the total rates from the $1p1h$ channel 
 go from $\Gamma/\Gamma^{free}_{\Lambda}=0.77$ in $^{12}_{\Lambda}$C to
1.3 in $^{208}_{\Lambda}$Pb and the ratios $\Gamma_{n}/\Gamma_{p}$ are all of
them of about $\Gamma_{n}/\Gamma_{p}\sim 0.53$.

We have also made calculations with the couplings for $K$ exchange from 
 \cite{Schaffner-Bielich:2000sy} and \cite{Savage:1996ca}. The qualitative
 results do not change much. In both cases they lead to smaller ratios and
 larger widths, particularly in the case of \cite{Schaffner-Bielich:2000sy},
 because the strength of the $K$ exchange is smaller and so are the interference
 terms. However, we should mention that  a fit with only three independent terms
 is done in \cite{Schaffner-Bielich:2000sy} while there is more freedom, 
 according to \cite{bookvertex}, where five independent terms contributing to
 the parity conserving part can be found.
 
 We can also make estimates of possible contributions from heavier mesons like
 the $\rho$ and the $K^*$. Given the small range of these exchanges, they can be
 easily accounted for by means of moderate changes in the $g'_{\Lambda}$
 parameters of $\pi$ and $K$ exchange. This is a usual way to account for $\rho$
 effects in the effective spin-isospin strong interaction via the Landau-Migdal
 effective force. We have checked that by increasing the $g'_{\Lambda}$
 parameter by $20\%$ in the case of the pion or the kaon exchange, the changes
 in both the $\Gamma_{n}/\Gamma_{p}$ ratio and the rates were smaller than
 $5\%$.

\begin{figure}[thb]
\centerline{\includegraphics[width=0.8\textwidth]{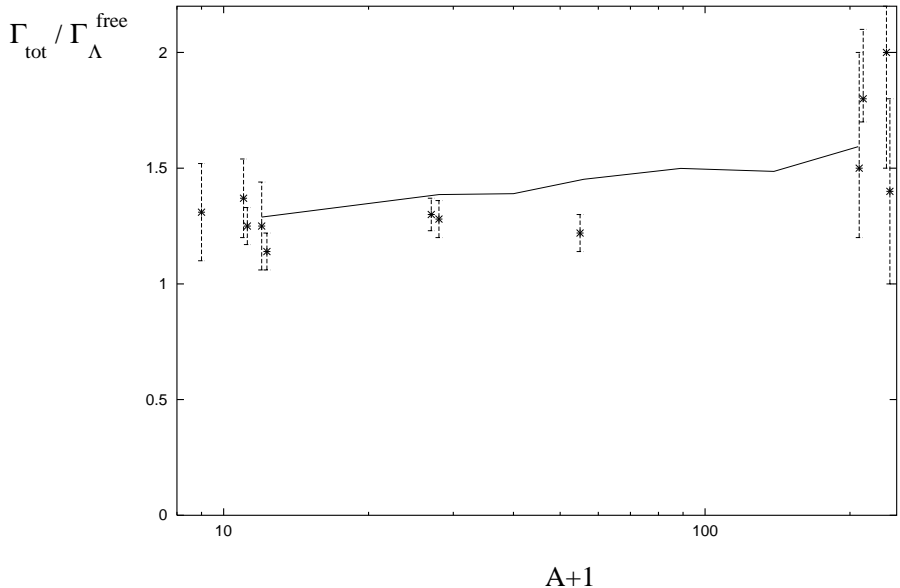}}
\caption{Calculated total $\Lambda$ decay widths versus experimental
data from table~\ref{park}. }
\label{fig:comptot}
\end{figure}

If we want to compare these results with experimental data we should still add
the mesonic contribution and the $2p2h$ induced one. For the mesonic
contribution we take the results from~\cite{Nieves:1993pm}, which agree well with
experiment in the measured cases. The mesonic rates are only relevant for the
lighter nuclei. We take $\Gamma_{M}=0.25 \textrm{ }\Gamma^{free}_{\Lambda}$ for $^{12}_{\Lambda}$C,
0.07 for $^{26}_{\Lambda}$Si and 0.03 for $^{40}_{\Lambda}$Ca and neglect this
contribution for heavier nuclei. The $2p2h$ induced contribution calculated
in~\cite{Ramos:1994xy} is $0.27 \textrm{ }\Gamma^{free}_{\Lambda}$ for $^{12}_{\Lambda}$C and 0.30 for the rest
of the nuclei. With these results we compute the total rates which we show in
table~\ref{final}. We have recalculated this contribution with the new form factor
and with the new values of $g^{\prime}$ that we use here, rescaling the $C_{0}$
parameter of \cite{Ramos:1997ik} in order to obtain the same strength for the
p-wave part of the pionic atoms optical potential. We obtain the same results as
in ref. \cite{Ramos:1997ik}, within $5\%$. The present status of the width measurements can be found in
table~\ref{park}. We can see that our total rates are 
% about $15\%$ bigger than
compatible with
the experimental numbers in the best measured nuclei. In heavy nuclei the
experimental errors are larger and our results are compatible with the
experiment.
In fig.~\ref{fig:comptot} we plot our results versus experiment for the
total $\Lambda$ decay width.

As for the $\Gamma_{n}/\Gamma_{p}$ ratios our results are compatible within 
 the experimental errors, on the lower side. However, one word of caution is necessary
here. The experimental analyses were done neglecting the $2p2h$ induced channel,
but it was observed in~\cite{Ramos:1994xy} that the inclusion of this mechanism in the
analyses of the data leads to different values of $\Gamma_{n}/\Gamma_{p}$. A
formula was given in this reference to correct the results of the old analysis
due to the consideration of this induced mechanism, but it assumed that all
particles were detected. The formula was corrected in \cite{Gal:1995qe} assuming
 that the
slow particles (with energies smaller than about 40 MeV) were not detected.
Detailed calculations of the spectra of protons and neutrons from the non-mesonic
decay were done in~\cite{Ramos:1997ik} but assuming a ratio of $1p1h$ to $2p2h$ induced
strength given by the OPE model alone, which as shown here overcounts the $1p1h$
strength. In view of this we just take the formula of~\cite{Gal:1995qe} and use it to
recalculate the experimental bands. The present bands for $^{12}_{\Lambda}$C are:
$1.33^{+1.12}_{-0.81}$\cite{Szymanski:1991ik}, $1.87^{+0.91}_{-1.59}$
\cite{Noumi:1995yd}, $0.70\pm
0.30$\cite{montwill}, $0.52\pm 0.16$\cite{montwill}. The lower bounds 
are 0.52, 0.29, 0.4, 0.36 respectively.
However, if we use the formula of~\cite{Gal:1995qe} assuming $\Gamma_{2p2h}/\Gamma_{nm}$
of the order of 0.3 one reduces the lower bound to values 0.2, 0.14, 0.1, 0.01
and the value obtained by us is well within present experimental ranges.

\section{Conclusions}

We have evaluated the non-mesonic proton and neutron induced $\Lambda$ decay
rates in nuclei, by including one pion, one kaon, correlated 
and uncorrelated two-pion exchange and $\omega$ exchange. We found
that the contribution of $K$ 
exchange was essential to reduce the total decay rate from the OPE results and
simultaneously increase the value of the $\Gamma_{n}/\Gamma_{p}$ ratio from
values around 0.12 for the OPE to values around 0.52. We also included the
$\sigma$ and uncorrelated two pion exchange and we found some cancellations
between them, such that the total contribution of the two-pion exchange to the
total rate and the $\Gamma_{n}/\Gamma_{p}$ ratio was small. 
% For revised version
Additional inclusion of the $\omega$ exchange made the
scalar-isoscalar contribution smaller, and while changing the
$\Gamma_{n}/\Gamma_{p}$ ratios only from 0.54 to 0.53, it helped a bit
in getting the total rates in better agreement with the data.
However, in these
results it was very important that the contribution of the $\sigma$ exchange had
opposite sign to the conventional contributions taking only the exchange of a
$\sigma$ particle. This change of sign was due to the presence
of the Adler zero in the scalar-isoscalar $\pi \pi$ interaction which makes the
amplitude change sign below $s=\mu^{2}/2$, which is the case here where we
have $s$ negative. We have also evaluated a weak parity violating $\sigma$
exchange term which we found negligible compared to the parity conserving one.

The total rates obtained are 
%fair, about $15\%$ larger than experiment as an average. 
good compared to the present data.
The ratios $\Gamma_{n}/\Gamma_{p}$ are considerably improved with
respect to the OPE ones and compatible with present experiments. We have also
 seen that, once the present experimental
data are corrected to account for the $2p2h$ channel the value of 0.53 obtained
here for the $\Gamma_{n}/\Gamma_{p}$ ratio is well within the present
experimental boundaries.

Further studies to evaluate the contribution from shorter distances, using for
instance quark models, like in~\cite{Sasaki:2000vi,oka2} would be most welcome.
 They
seem to lead to even larger values of $\Gamma_{n}/\Gamma_{p}$ but also larger
total decay rates. In any case our study of the long and intermediate distances
has shown that one can obtain values for $\Gamma_{n}/\Gamma_{p}$ considerably larger than those
given by the OPE model while still having total rates in good
agreement with experiment.

\subsection*{Note added in proof}
    While our paper was in print another related paper appeared in the
web \cite{new} paying especial attention to the final state interaction
(FSI) of the two nucleons after the $\Lambda$ decay. Tables IV and V of
that paper show  the effects of the FSI, but the initial state
interaction (ISI) is already accounted for. One should not compare these
results  with ours, since our correlation function is meant to account
for initial as well as final state correlations. Comparison of our results 
for one pion exchange and $^{12}_\Lambda$C  with those of \cite{new}, complemented by 
the effects of ISI found in \cite{PRB97} is possible. Depending on the 
strong potential used, the combined effect of ISI and FSI in \cite{new},
\cite{PRB97} is a reduction by a factor ranging for  1.4 to 1.9.  
Our correlation function gives us a reduction of a factor 1.5.

\subsection*{Acknowledgements}
We wish to acknowledge multiple and useful discussions with A. Ramos and A.
Parre\~no. Useful information from A. Gal and M. Oka is much appreciated. 
 We would also like to acknowledge financial
support from the DGICYT under contracts PB96-0753, PB98-1247 and
AEN97-1693, from the Generalitat de Catalunya under grant
SGR98-11
and from the EU TMR network Eurodaphne, contract no.
ERBFMRX-CT98-0169. J.E.P. thanks finantial support from the Ministerio
 de Educaci\'on y Cultura. The work by D.J. is supported by the
Spanish Ministry of Education in the Program ``Estancias de Doctores y
Tecn\'ologos Extranjeros en Espa\~{n}a''. 

\section*{Appendix}

In this Appendix we briefly review how to implement the short range
correlation in the effective interactions of $\Lambda N
\rightarrow NN$ and $NN \rightarrow NN$. We follow here closely the steps of
\cite{Oset:1979bi} for this purpose. The hard core in the short
range is produced by a strong repulsive force independent on
spin and isospin in the $NN$ interaction, and is  very well approximated
by a local correlation function $g(r)$. 
Then, we may write the effective interaction
$G_{NN}$ through
one meson exchange potential $V_{NN \rightarrow NN}$ as
\begin{equation}
   G_{NN}(r) = g(r) V_{NN} (r) \ ,
\end{equation}
where typically  $g(r)$ vanishes as $r\rightarrow 0$ and goes to $1$ as $r
\rightarrow \infty$. One of the practical choices of $g(r)$ is 
\begin{equation}
   g(r) = 1 - j_0(q_c r) \ . \label{corr}
\end{equation}
The analysis of the $NN$ interaction suggested that  eq. (\ref{corr}) with 
$q_c = 780 {\rm MeV}/c$ gives a fair correlation
function \cite{Oset:1979bi}. 

With eq. (\ref{corr}) the effective interaction in momentum space is
given by
\begin{equation}
G_{NN}(q) = [V_l(q)\hat{q}_i \hat{q}_j + V_t(q)(
    \delta_{ij}-\hat{q}_i \hat{q}_j)] \sigma_i^{(1)} \sigma_j^{(2)}
    \vec{\tau}^{(1)} \cdot \vec{\tau}^{(2)}
\end{equation}
with
\begin{eqnarray}
 V_l(q) &=& {f_{\pi NN}^2 \over \mpi^2} [\vec{q}^{\, 2} D_\pi(q)F_\pi^2(q) + g_l(q)] \\
 V_t(q) &=& {f_{\pi NN}^2 \over \mpi^2} [\vec{q}^{\, 2} D_\rho(q)F_\rho^2(q)
 C_\rho  + g_t(q)]
\end{eqnarray}
The short range correlations are taken into account in the $g_l(q)$
and $g_t(q)$:
\begin{eqnarray}
  g_l(q) &=&  - \left( \vec{q}^{\, 2} + {q_c^2 \over 3} \right) 
 \tilde{F}_\pi^2(q) \tilde{D}_\pi(q) - {2 q_c^2 \over 3} C_\rho
 \tilde{F}_\rho^2(q) \tilde{D}_\rho(q) \label{gl} \ , \\
  g_t(q) &=& - {q_c^2 \over 3} \tilde{F}_\pi^2(q) \tilde{D}_\pi(q)
   -\left(\vec{q}^{\, 2} + {2 q_c^2 \over 3}\right) C_\rho
 \tilde{F}_\rho^2(q) \tilde{D}_\rho(q) \label{gt} \ .
\end{eqnarray}
The tilde on the functions means that $\vec q^{\textrm{ }2}$ must be changed by 
$\vec q^{\textrm{ }2} + q_c^2$ in the argument of the function.
$C_\rho$ is the ratio of the $\rho NN$ coupling
to the $\pi NN$ coupling:
$  C_\rho = (f_\rho/m_\rho)/(f_{\pi NN}/\mpi) = 2 $. Here we rescale these functions to keep consistency with the
spin-isospin effective nuclear force:
\begin{equation}
    g_{l,t}(q) \rightarrow g^\prime {g_{l,t}(q) \over g_{l,t}(0)}
\end{equation}
with $g^\prime = 0.7$. 
$D_\pi(q)$ and $D_\rho(q)$ are the propagators of $\pi$ and
$\rho$, respectively, and $F_\pi(q)$ and $F_\rho(q)$ denote the form
factor for $\pi NN$ and $\rho NN$, which are given by
\begin{eqnarray}
   F_\pi(q) &=& {\Lambda_\pi^2 \over \Lambda_\pi^2 - q^2} \\
  F_\rho(q) & =& { \Lambda_\rho^2 - m_\rho^2 \over \Lambda^2_{\rho} - q^2} 
\end{eqnarray}
with $\Lambda_\pi = 1.0$ GeV and $\Lambda_\rho = 2.5$ GeV. 

Similarly we introduce the short range correlation into the effective
interaction $\Lambda N \rightarrow NN$:
\begin{equation}
   G_{\Lambda N \rightarrow NN}(r) = g(r) V_{\Lambda N \rightarrow
   NN} (r) 
\end{equation}
where $V_{\Lambda N \rightarrow NN}$ is the potential with 
one $\pi$ and $K$ exchanges in this case.
We use the same correlation function and $q_c$ for $G_{\Lambda N
\rightarrow NN}(r)$ and for $G_{NN}(r)$, as usually done.

Then the effective interaction with one $\pi$ exchange 
for the parity violating part is written as
\begin{equation}
   G_{\Lambda N \rightarrow NN}^{\pi, {\rm s-wave}}(q) =
   V_s^\prime(q)\ 
   \hat{q}_i \ \sigma_i^{(2)} \ \vec\tau^{(1)} \cdot \vec\tau^{(2)}  
\end{equation}
while the parity conserving part is split into the longitudinal and
transverse components: 
\begin{equation}
   G_{\Lambda N \rightarrow NN}^{\pi, {\rm p-wave}}(q) =
    [V_l^\prime(q)\hat{q}_i \hat{q}_j
   + V_t^\prime(q)(\delta_{ij}-\hat{q}_i \hat{q}_j)] 
    \sigma_i^{(1)} \sigma_j^{(2)} \vec{\tau}^{(1)} \cdot \vec{\tau}^{(2)} 
\end{equation}
with
\begin{eqnarray}
 V_l^\prime(q) &=& {f_{\pi NN} \over \mpi}{P \over \mpi} [\vec{q}^{\, 2} 
  D_\pi(q)F_\pi^2(q) + g_l^\Lambda(q)] \\
 V_t^\prime(q) &=& {f_{\pi NN} \over \mpi} {P \over \mpi} g_t^\Lambda(q) \\
 V_s^\prime(q) &=& {f _{\pi NN}\over \mpi} S [
  D_\pi(q)F_\pi^2(q) +g_s^\Lambda(q)]|\vec{q}| 
\end{eqnarray}
The form factor is assumed to be the same as for the strong $\pi NN$ vertex. 
The short range correlations are considered in $g^\Lambda_i$:
\begin{eqnarray}
 g_l^\Lambda(q) &=& - \left( \vec{q}^{\, 2} + {q_c^2 \over 3} \right) 
 \tilde{F}_\pi^2(q) \tilde{D}_\pi(q) \\
 g_t^\Lambda(q) &=& - {q_c^2 \over 3} \tilde{F}_\pi^2(q) \tilde{D}_\pi(q)
 \\
 g_s^\Lambda(q) &=& - \tilde{F}_\pi^2(q) \tilde{D}_\pi(q)
\end{eqnarray}
where the tilde is defined as for the case of the $NN$ interaction.

Similarly, the effective interaction with one  kaon exchange are
given as
\begin{equation}
   G_{\Lambda N \rightarrow NN}^{K, {\rm p-wave}}(q) =
    [V_l^{\prime, K}(q)\hat{q}_i \hat{q}_j
   + V_t^{\prime, K}(q)(\delta_{ij}-\hat{q}_i \hat{q}_j)] 
    \sigma_i^{(1)} \sigma_j^{(2)} C^{K,N}_P
\end{equation}
\begin{equation}
   G_{\Lambda N \rightarrow NN}^{K, {\rm s-wave}}(q) =
   V_{s,K}^\prime(q)\  \hat{q}_i\  \sigma_i^{(1)}\  C^{K,N}_S 
\end{equation}
with 
\begin{eqnarray}
 V_{l,K}^\prime(q) &=& -{f_{K\Lambda N} \over \mpi}
     {P \over \mpi} [\vec{q}^{\, 2} D_K(q)F_K^2(q) + g_{l,K}^\Lambda(q)] \\
 V_{t,K}^\prime(q) &=& -{f_{K\Lambda N} \over \mpi} 
   {P \over \mpi} g_{t,K}^\Lambda(q) \\
 V_{s,K}^\prime(q) &=& -{f_{K\Lambda N} \over \mpi} S [ 
  D_K(q)F_K^2(q)  + g_{s,K}^\Lambda(q)]|\vec{q}| 
\end{eqnarray}
Here $D_K(q)$ is the $K$ propagator and $F_K(q)$ denotes the form
factor of the $K \Lambda N$. We use the same form factor for the weak
$KNN$ vertex:
\begin{equation}
F_K(q) = {\Lambda_K^2 \over \Lambda_K^2 -q^2}
\end{equation}
with $\Lambda_K = 1.0$ GeV.
The short range correlations are accounted for by means of the $g\prime$ coefficients

\begin{eqnarray}
 g_{l,K}^\Lambda(q) &=& - \left( \vec{q}^{\, 2} + {q_c^2 \over 3} \right) 
 \tilde{F}_K^2(q) \tilde{D}_K(q) \\
 g_{t,K}^\Lambda(q) &=& - {q_c^2 \over 3} \tilde{F}_K^2(q)
 \tilde{D}_K(q) \\
 g_{s,K}^\Lambda &=& - \tilde{F}_K^2(q) \tilde{D}_K(q)
\end{eqnarray}

\end{document}